\documentclass{article}

\usepackage[nonatbib,preprint]{neurips_2022}

\usepackage[utf8]{inputenc} 
\usepackage[T1]{fontenc}    
\usepackage{hyperref}       
\usepackage{url}            
\usepackage{booktabs}       
\usepackage{amsfonts}       
\usepackage{nicefrac}       
\usepackage{microtype}      
\usepackage{xcolor}         
\usepackage{tabto}    
\usepackage[superscript,biblabel]{cite}
\usepackage{setspace}  
\usepackage{titlesec}

\usepackage{subfig}
\usepackage{wrapfig}
\usepackage{tikz}
\usepackage{tikz-cd}
\usetikzlibrary{cd}
\usepackage{microtype}
\usepackage{graphicx}
\usepackage{booktabs} 
\usepackage{pifont}       
\usepackage{bbding}       %
\usepackage{fontawesome}  
\usepackage{jabbrv}

\usepackage{amsmath}
\usepackage{amssymb}
\usepackage{mathtools}
\usepackage{amsthm}
\usepackage{arydshln}
\usepackage[capitalize,noabbrev]{cleveref}

\usepackage{mathtools} 

\usepackage{amsmath,amsfonts,bm}

\def\1{\bm{1}}

\def\rd{{\textnormal{d}}}

\DeclareMathAlphabet{\mathsfit}{\encodingdefault}{\sfdefault}{m}{sl}
\SetMathAlphabet{\mathsfit}{bold}{\encodingdefault}{\sfdefault}{bx}{n}

\def\calP{{\mathcal{P}}}

\def\calX{{\mathcal{X}}}
\def\calY{{\mathcal{Y}}}

\newcommand{\R}{\mathbb{R}}

\newcommand{\fracpartial}[2]{\frac{\partial #1}{\partial  #2}}
\newcommand{\fracdiff}[2]{\frac{\rd #1}{\rd  #2}}

\newcommand{\norm}[1]{\lVert#1\rVert}

\makeatletter
\renewcommand{\fnum@figure}{\textbf{Fig. \thefigure} | }
\renewcommand{\fnum@table}{\textbf{Table \thetable \ |}}

\makeatother
\usepackage[labelsep=space]{caption}
\usepackage[
singlelinecheck=false 
]{caption}

\usepackage{authblk}

\title{
Optimal Transport for Generating Transition States in Chemical Reactions
}

\author[1, 2, 3, $ \dag $, *]{Chenru Duan}
\author[4, $ \dag $, *]{Guan-Horng Liu}
\author[5, $ \dag $, *]{Yuanqi Du}
\author[6]{Tianrong Chen}
\author[3, 7, 8]{Qiyuan Zhao}
\author[1, 2]{Haojun Jia}
\author[5]{Carla P. Gomes}
\author[4]{Evangelos A. Theodorou}
\author[1, 2]{Heather J. Kulik}
\affil[1]{Department of Chemistry, Massachusetts Institute of Technology, Cambridge, MA, 02139}
\affil[2]{Department of Chemical Engineering, Massachusetts Institute of Technology, Cambridge, MA, 02139}
\affil[3]{Deep Principle, Inc., Cambridge, MA, 02139}
\affil[4]{School of Aerospace Engineering, Georgia Institute of Technology, Atlanta, GA, 30332}
\affil[5]{Department of Computer Science, Cornell University, Ithaca, NY, 14850}
\affil[6]{School of Electrical and Computer Engineering, Atlanta, GA, 30332}
\affil[7]{Department of Medicinal Chemistry, College of Pharmacy, University of Michigan, Ann Arbor, MI 48109}
\affil[8]{Department of Chemistry, University of Michigan, Ann Arbor, MI, 48109}
\affil[$ \dag $]{These authors contribute equally}
\affil[*]{Corresponding to: duanchenru@gmail.com, guanhorng.liu@gmail.com, yuanqidu@cs.cornell.edu}

\begin{document}

\maketitle

\begin{abstract}
Transition states (TSs) are transient structures that are key in understanding reaction mechanisms and designing catalysts but challenging to be captured in experiments.
Alternatively, many optimization algorithms have been developed to search for TSs computationally.
Yet the cost of these algorithms driven by quantum chemistry methods (usually density functional theory) is still high, posing challenges for their applications in building large reaction networks for reaction exploration.
Here we developed React-OT, an optimal transport approach for generating unique TS structures from reactants and products.
React-OT generates highly accurate TS structures with a median structural root mean square deviation (RMSD) of 0.053Å and median barrier height error of 1.06 kcal/mol requiring only 0.4 second per reaction.
The RMSD and barrier height error is further improved by roughly 25\% through pretraining React-OT on a large reaction dataset obtained with a lower level of theory, GFN2-xTB.
We envision that the remarkable accuracy and rapid inference of React-OT will be highly useful when integrated with the current high-throughput TS search workflow. 
This integration will facilitate the exploration of chemical reactions with unknown mechanisms.

\end{abstract}

\setstretch{1.8}

\section*{Introduction}
Transition states (TSs) are central in understanding the kinetics and mechanisms of chemical reactions\cite{TSTReview,EEricRev}.
Accurate TS structures reveal precise elementary reaction steps on the potential energy surface (PES), enabling the construction of large reaction networks for complex chemical reactions \cite{ZimmermanWIREsRev,ReactNetworkAnnualRev} and the design of new catalysts\cite{Grzybowski2024,WalshDD2024,KulikChemRev,NvJustinNatChem}.
However, since they  have higher free energy compared to reactants and products on reaction pathways and are affected by dynamic processes, TSs are transient structures that often live at a time scale of femtoseconds, which makes them impossible to be isolated and characterized experimentally. 
Very few studies have successfully unraveled elusive TS properties or their structures in experiments. 
Ultrafast millimeter-wave vibrational spectra, for example, was used to analyze characteristic patterns in isomerizing systems where their TS energies and properties are extracted from frequency-domain data\cite{RWField2020}.
More recently, ultrafast electron diffraction was used to get a more direct observation of TS structures for a photochemical electrocyclic ring-opening reaction\cite{Wolf2023}.
These experimental techniques, however, are expensive and cannot be universally applied in all types of reactions.

\qquad In parallel, defined as a first-order saddle point on the PES, the TS of any chemical reaction can be searched systemically by a suite of optimization algorithms coupled with quantum chemistry calculations (for example, density functional theory [DFT]\cite{DFTReview}).
These TS search algorithms have been developed over the past twenty years, with salient examples being string methods\cite{string_method}, growing string methods\cite{growing_string}, nudged elastic band (NEB) methods,\cite{NEB} artificial force induced reaction (AFIR),\cite{maeda2014AFIR} and stochastic surface walking methods (SSWM)\cite{shang2013SSWM}. 
Together with these algorithms, comprehensive reaction networks\cite{ReactNetworkAnnualRev,Durant1996} can be constructed by either iteratively enumerating potential elementary reactions on-the-fly\cite{zimmerman2013ZStucture,Reiher2018,ReiherMRNetwork,QiyuanNCS2021}, exploration PES starting from a local minimum\cite{shang2013SSWM,maeda2014AFIR}, or propagating biased \textit{ab initio} molecular dynamics for enhanced sampling\cite{Nanoreactor,NA-Nanoreactor,Zeng2020}.
For a reasonable-sized reaction network, however, thousands of TS structures need to be optimized, requiring millions of DFT single-point calculations\cite{KinBot,vonLilienfeld2020, Margraf2023}.
The large number of chemical species involved in reaction networks call for the need to reducing the computational cost of TS search for a single elementary step in chemical reactions.

\qquad Recently, machine learning (ML) has demonstrated promise for accelerating the search of TSs.
The ideas include developing ML potential as a surrogate for DFT during TS optimizations (for example, in NEB)\cite{NeuralNEB, ANI-1xnr} and formulating the transition path sampling problem as a "shooting game" solved by reinforcement learning \cite{RLTS}. 
It has also been formulated as a stochastic optimal control problem to learn a controlled stochastic process from the reaction to the product\cite{pips}.
The TS search problem can also be viewed as a 3D structure generation task, which can then be addressed by equivariant graph neural networks~\cite{GreenPCCP,EquiReact}, generative-adversarial networks\cite{TSGAN}, a combination of gated recurrent neural networks and transformers\cite{ChoiNatComm}, or denoising diffusion probabilistic models\cite{ddpm,2DTSDiff,OAReactDiff,AlanDiffusion2024}.
Our prior work, OA-ReactDiff\cite{OAReactDiff}, leverages a diffusion model for elementary chemical reactions, directly generating a set of reactants, TS structure, and products jointly.
By preserving all required symmetries of chemical reactions, OA-ReactDiff achieves the state-of-the-art performance on generated structure similarity. 
Despite its high accuracy, the stochastic nature of OA-ReactDiff requires multiple runs of sampling for a reaction of interest and the use of a ranking model\cite{DFARec,DiffDock} to recommend one out of many generated TSs.
This workflow leads to both an additional cost of sampling and unfavored randomness in a double-ended TS search with reactants and products provided, which would be troublesome in its practical usage when replacing or reducing DFT calculations in high throughput TS search.

\qquad In this work, we developed React-OT, an \underline{o}ptimal \underline{t}ransport approach to generate TSs of an elementary \underline{react}ion in a fully deterministic manner. 
Compared to OA-ReactDiff, React-OT eliminates the need of training an additional ranking model and reduces the number of inference evaluations of the denoising model from 40,000 to 50, achieving a near 1000-fold acceleration. 
With React-OT, highly accurate TS structures can be deterministically generated in 0.4 seconds.
In addition, React-OT outperforms OA-ReactDiff in both structural similarity and barrier height estimation by 30\%.
When pretrained on RGD1-xTB\cite{RGD1}, a dataset containing 760,615 elementary reactions computed with low-cost semi-empirical quantum chemistry methods (i.e., GFN2-xTB\cite{GFN2-xTB}), React-OT improves further by 25\%, reaching a median RMSD of 0.044 Å and median error of 0.74 kcal/mol in barrier height prediction.
To accelerate the TS search, we replace DFT-optimized reactant and product geometries with those optimized using much cheaper GFN2-xTB. 
In this scenario, React-OT continues to exhibit comparable performance, achieving a median RMSD of 0.049 Å and median error of 0.79 kcal/mol in barrier height calculations.
Furthermore, we integrated React-OT in the high throughput DFT-based TS optimization workflow, where an uncertainty quantification model is used to activate DFT-based TS search only when the generated TS from React-OT is uncertain.
With this new workflow, chemical accuracy can be achieved on the generated TS structures using just one seventh of the computational resources required for full reliance on DFT-based TS optimizations.
The high quality of generated TS, the extremely low cost in sampling, and the simplicity of its integration to high throughput computational workflow characterize React-OT as a promising model in constructing large reaction networks for studying new chemical reactions with unexplored mechanisms.

\section*{Results}

\begin{figure*}[t!]
    \centering
    \includegraphics[width=0.98\textwidth]{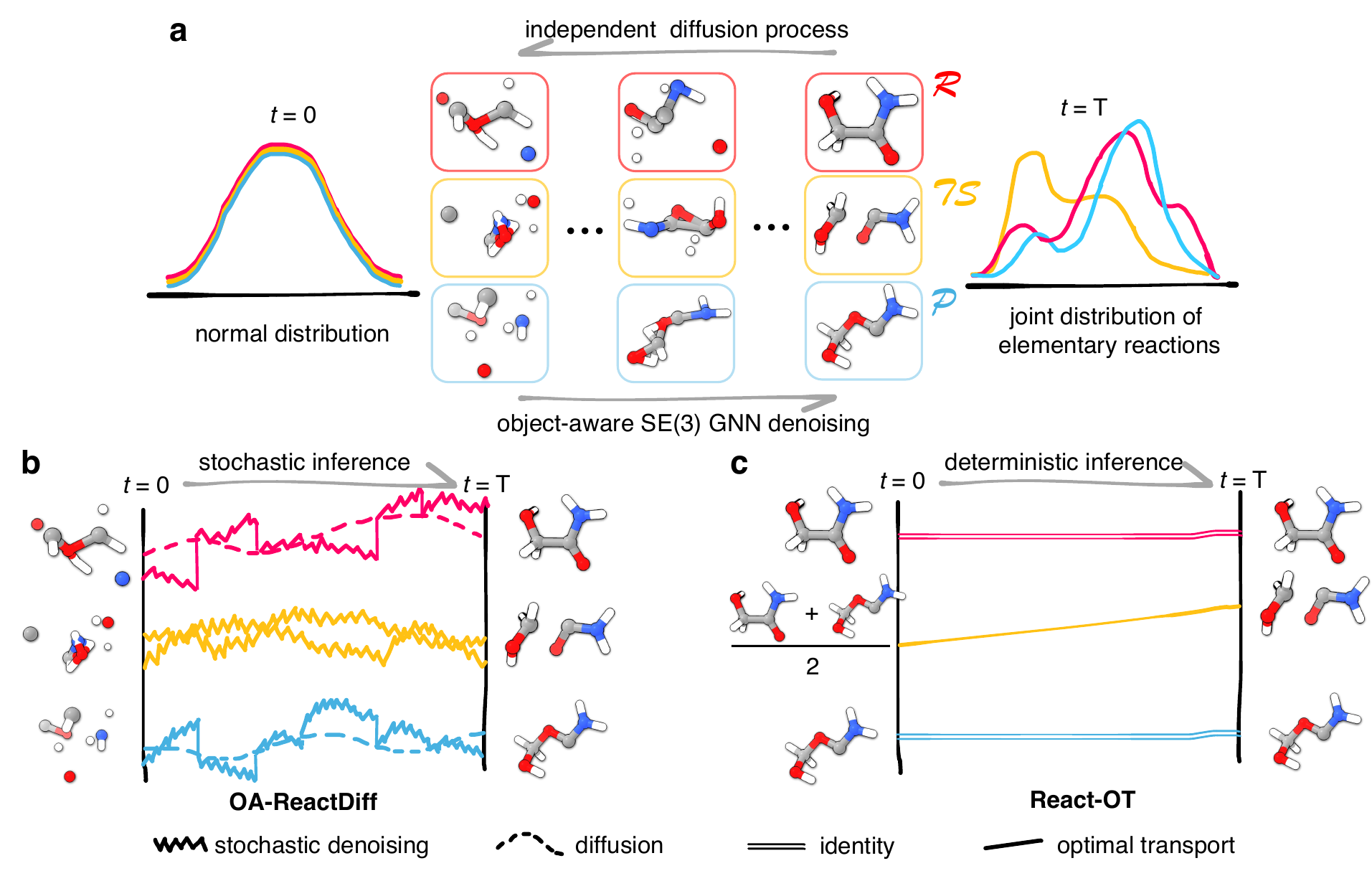}
    \caption{\textbf{Overview of the diffusion model and optimal transport framework for generating TS.}
    \textbf{a.} Learning the joint distribution of structures in elementary reactions (reactant in red, TS in yellow, and product in blue). A forward diffusion process brings the joint distribution at $t=T$ to independent normal distribution at $t=0$. Backward, an object-aware SE(3) GNN is trained with denoising objective to recover the normal distribution to the original joint distribution.
    \textbf{b.} Stochastic inference with inpainting in OA-ReactDiff. Starting with samples drawn from normal distribution, the trained GNN is applied to denoise the reactant, TS, and product. A diffusion process on reactant and product is combined with the denoising process to ensure the end-point reactant and product at $t=T$ are the same as true reactant and product.
    \textbf{c.} Deterministic inference with React-OT. Both the reactant and product are unchanged throughout the entire process from $t=0$ to $t=T$. The linear interpolation of reactant and product is provided as the initial guess structure at $t=0$, followed by optimal (i.e., linear) transport to the final TS.
    Atoms are colored as follows: C for gray; N for blue, O for red, and H for white.
    }
    \label{fig:overview}
    
\end{figure*}

\paragraph{Overview of React-OT.}
A diffusion model learns the underlying distribution of observed samples through training a scoring network by a denoising objective \cite{diffusion2015, ddpm, scoresde} (see \textit{\nameref{diffusion}}).
Our earlier work, OA-ReactDiff\cite{OAReactDiff}, which satisfies all the symmetries in elementary reactions, essentially learns the joint distribution of paired reactants, TSs, and products (Fig. \ref{fig:overview}a, Supplementary Text \ref{Supp:required_symmetries}).
With the learned joint distribution, one can either generate a new reaction from scratch or only generate a TS structure conditioned on fixed reactants and products (RP), resembling the setup of the double-ended TS search problem (Fig. \ref{fig:overview}b, see \textit{\nameref{inpainting}}).
The standard Gaussian distribution, which is far away from a reasonable guess of the TS structure, was used as the starting point for sampling (Supplementary Figure \ref{Supp:rmsd_guss}).
The fact that initial structures at $t=0$ are randomly sampled from a Gaussian also leads to inevitable stochasticity in the final generated TS structures (Fig. \ref{fig:overview}b).

\qquad To address this challenge, we reformulate the double-ended TS search problem in a dynamic transport setting and utilize an objective resembling flow matching\cite{FlowMatching} to achieve the optimal transport in reactions\cite{I2SB,somnath2023aligned} (see \textit{\nameref{ot}}).
For simplicity, we use the linear interpolation of RP as the guess starting point and keep the RP constant during the transport process (Fig. \ref{fig:overview}c).
This way, we avoid any stochastic processes during the sampling phase.
React-OT also uses a transition kernel that is object-aware SE(3) equivariant, thus fulfilling all symmetries required in modeling an elementary reaction (see \textit{\nameref{equivariance}}).
In essence, the key benefit of React-OT are three folds:
First, it simulates the sampling process as an ordinary differential equation instead of as a stochastic differential equation in diffusion models.
Therefore, the generated TS with React-OT is deterministic, in line with the fact that there is only one unique TS structure given the paired RP conformations\cite{zhao2022RCS,sindhu2019theoretical,koda2024locating}.
Second, utilizing a relatively reasonable initial guess and pushing the sampling path closer to optimal transport, React-OT generates TS structures with higher accuracy at lower cost.
Third, with fully deterministic inference, React-OT only needs to be run once to generate the final TS structure, greatly simplifying its application in realistic computational workflows.
These advantages make React-OT a model that is trustworthy and extremely low-cost, adequately starting to replace actual DFT calculations in high throughput computing.

\begin{figure*}[t!]
    \centering
    \includegraphics[width=0.98\textwidth]{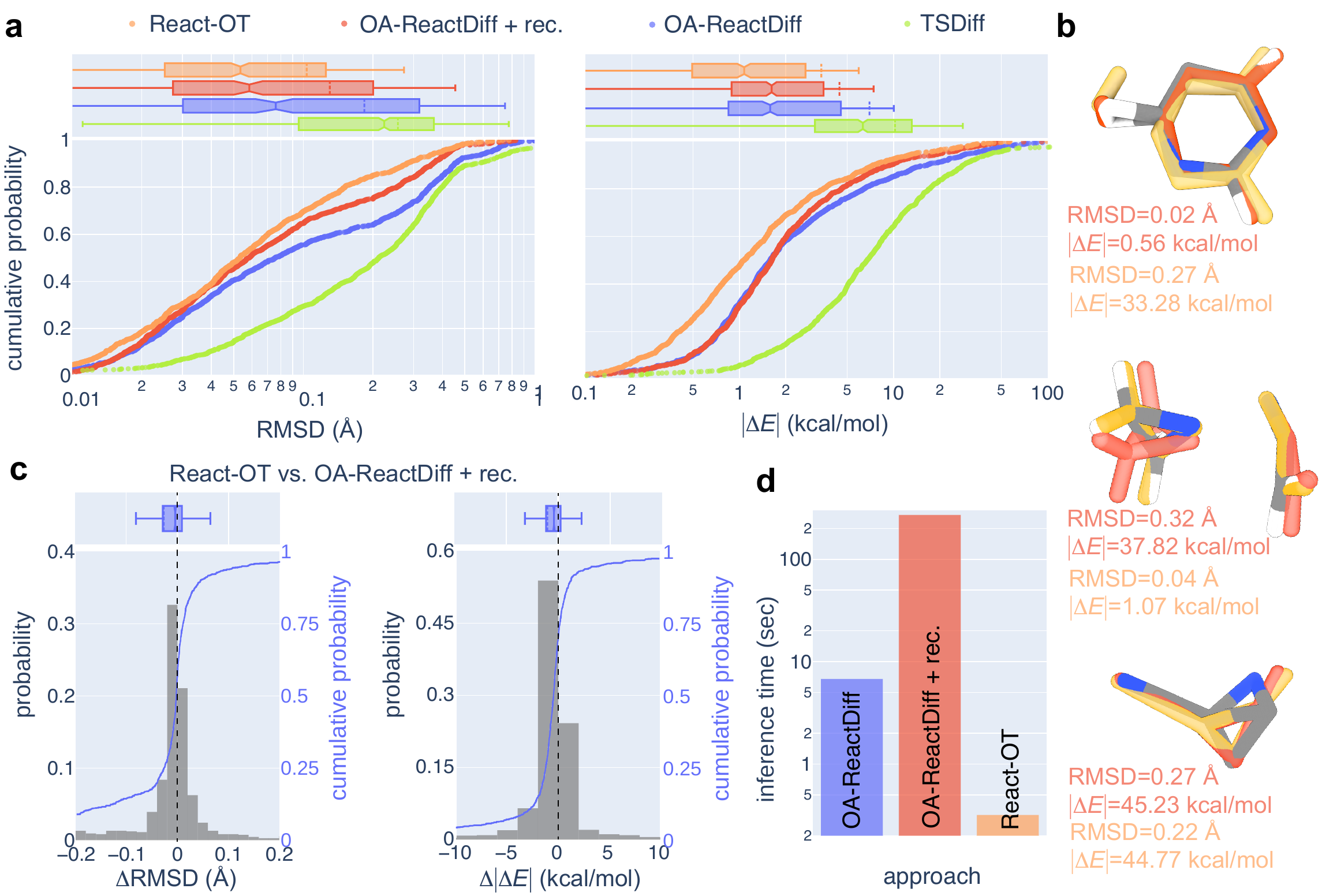}
    \caption{\textbf{Structural and energetic performance of diffusion and optimal transport generated TS structures.}
    \textbf{a.} Cumulative probability for structure root mean square deviation (RMSD) (left) and absolute energy error ($|\Delta \mathit{E}_\mathrm{TS}|$) (right) between the true and generated TS on 1,073 set-aside test reactions. single-shot OA-ReactDiff\cite{OAReactDiff} (blue), 40-shot OA-ReactDiff with recommender (red), single-shot TSDiff\cite{2DTSDiff} (green), and React-OT TS (orange) are shown. Both RMSD and $|\Delta \mathit{E}_\mathrm{TS}|$ are displaced in log scale for visibility of low error regime.
    \textbf{b.} Reference TS structure, OA-ReactDiff TS sample (red), and React-OT structure (orange) for select reactions. RMSD and $|\Delta \mathit{E}_\mathrm{TS}|$ for OA-ReactDiff and React-OT structures are shown in text with their corresponding color. Atoms in the reference TS are colored as follows: C for gray; N for blue, O for red, and H for white.
    \textbf{c.} Histogram (gray, left y axis) and cumulative probability (blue, right y axis) showing the difference of RMSD (left) and $|\Delta \mathit{E}_\mathrm{TS}|$ (right) between OA-ReactDiff recommended and React-OT structures compared to reference TS. Negative $\Delta$RMSD or $\Delta|\Delta \mathit{E}_\mathrm{TS}|$ suggests React-OT structure is of higher quality. A box plot (blue) for $\Delta$RMSD and $\Delta|\Delta \mathit{E}_\mathrm{TS}|$ is shown above the histogram, correspondingly.  A dashed vertical line is shown for no deviation between two structures.
    \textbf{d.} Inference time in seconds for single-shot OA-ReactDiff (blue), 40-shot OA-ReactDiff with recommender (red), and React-OT (orange). The y axis is displaced in log scale for visibility of the extremely low inference time for React-OT.
    }
    \label{fig:performance}
\end{figure*}

\begin{figure*}[t!]
    \centering
    \includegraphics[width=0.98\textwidth]{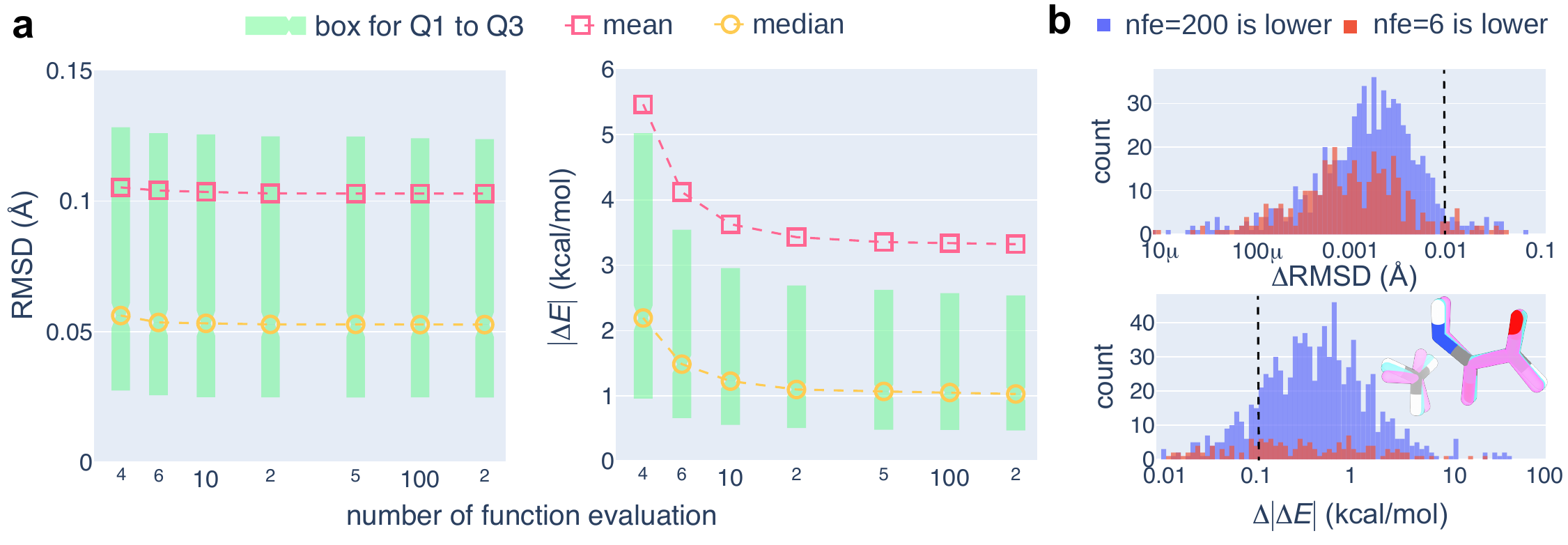}
    \caption{\textbf{Performance of React-OT with respect to the number of function evaluation (nfe).}
    \textbf{a.} Distribution of RMSD (left) and $|\Delta \mathit{E}_\mathrm{TS}|$ (right) between the true and generated TS on 1,073 set-aside test reactions, where the mean is shown in pink and median is shown in yellow. The first and third quarter is bounded by a green box.
    \textbf{b.} Absolute difference in RMSD (top) and $|\Delta \mathit{E}_\mathrm{TS}|$ (bottom) for generated structures at nfe=6 and nfe=200. Structures where nfe=200 gives better quality is shown in blue, otherwise shown in red. A threshold below which the comparison is not chemically meaningful is shown (dashed vertical line). Reference TS structure and React-OT generated structure at nfe=6 (pink) and nfe=200 (skyblue) for a select reaction. Atoms in the reference TS are colored as follows: C for gray; N for blue, O for red, and H for white.
    }
    \label{fig:nfe}
\end{figure*}

\vskip 0.1in
\paragraph{Generating accurate TS structure in 0.4 seconds.}
We trained React-OT on Transition1x\cite{ts1x}, a dataset that contains paired reactants, TSs, and products calculated from climbing-image NEB \cite{CINEB} obtained with DFT ($\omega$B97x/6-31G(d))\cite{wb97x,631gs}.
Transition1x contains 10,073 organic reactions with diverse reaction types sampled from enumeration \cite{Grambow2020, GreenDataApp} of 1,154 reactants in GDB7 \cite{gdb17} dataset, which has up to seven heavy atoms (C, N, and O) and 23 atoms in total. 
For the ease of comparison, we use the same train-test partitioning as in OA-ReactDiff, with 9,000 elementary reactions in training and leaving the remaining 1,073 unseen reactions as the test set (see \textit{\nameref{training}}).
In React-OT, the object-aware version of LEFTNet\cite{leftnet, OAReactDiff} is used as the scoring network to fit the transition kernel (see \textit{\nameref{leftnet}}).

\qquad React-OT achieves a mean RMSD of 0.103 Å between generated and true TS structures on the set-aside test reactions of Transition1x, significantly improved upon previous state-of-the-art results on ML generation for TS structures\cite{OAReactDiff,2DTSDiff}.
The mean structural RMSD from React-OT is halved compared to a diffusion model that only utilizes 2D graphs (TSDiff\cite{2DTSDiff}, 0.252 Å) and that uses 3D structures of RP explicitly (OA-ReactDiff\cite{OAReactDiff}, 0.180 Å).
React-OT also outperforms the combination of 40-shots sampling of OA-ReactDiff and a recommender by 26\%, which gives 0.130 Å mean RMSD at a 40x cost of one-shot OA-ReactDiff.
With React-OT, the likelihood of finding a TS below a certain RMSD is higher than that of both diffusion models, highlighting the excellent accuracy of React-OT across all test reactions regardless of their level of complexity (Fig. \ref{fig:performance}a).
Similarly, React-OT halves the mean absolute error (MAE) on barrier height estimation from 10.37 kcal/mol of single-shot TSDiff and 6.26 kcal/mol of single-shot OA-ReactDiff to 3.34 kcal/mol, irrespective to the actual barrier height of different reactions (Supplementary Figure \ref{Supp:abs_barrier_vs_barrier_error}).
Among 1073 reactions, 62\% of the TS structures generated by React-OT give a more precise TS structure compared to 40-shot OA-ReactDiff + recommender and 66\% for a more accurate barrier height (Fig. \ref{fig:performance}c, Supplementary Figure \ref{Supp:ot_vs_diff_best}).
It is found that the RMSD and barrier height error of the React-OT generated TS have good positive correlation, demonstrating the effectiveness of both metrics (Supplementary Figure \ref{Supp:rmsd_vs_barrier_error}).
In chemical reactions, one order of magnitude difference in reaction rate is often used as the characterization of chemical accuracy, which translates to 1.58 kcal/mol in barrier height error assuming a reaction temperature of 70°C\cite{ChemAccFu2022,ChemAccAdamo2022}.
With this definition, over 64\% of TS from React-OT falls in chemical accuracy, while the number is only 11\% for TSDiff and 49\% for OA-ReactDiff + recommender. 

\qquad The superior performance in both the structural similarity and barrier height estimation of React-OT can be ascribed to its more reasonable initial guess in combination with the optimal transport objective (Supplementary Table \ref{Supp:table_ablation_study}).
As a result, React-OT has much higher chance of obtaining the correct inter-molecular arrangement for TS that contains multiple fragments (Fig. \ref{fig:performance}b, middle).
With diffusion models, however, one can take advantage of their stochasticity and keep generating different conformations of TS until getting the most desired one\cite{2DTSDiff, OAReactDiff}.
Still, there are only very few reactions, in 40 runs of sampling, where OA-ReactDiff is able to generate much more accurate TS structures, with H atom transfer in 5H-pyrimidin-5-ide as an example (Fig. \ref{fig:performance}b, top).
Due to the limited size of training data, the parameterization of the scoring network is not perfect, leading to a few reactions where both React-OT and OA-ReactDiff generate the same TS structure that deviates significantly from the true TS (Fig. \ref{fig:performance}b, bottom).
These reactions mostly involve chemical species that are radical or those do not obey octet rule, which are rare in Transition1x and are thus difficult to learn (Supplementary Figure \ref{Supp:non_octect_examples}).
Starting from a more reasonable TS guess and trained towards optimal transport paths, React-OT only requires, on average, 0.39 seconds to generate a high-quality TS structure (Fig. \ref{fig:performance}d).
Interestingly, we do not observe a strong correlation between the quality of the initial guess structure and the RMSD for the React-OT generated structure, indicating the great effectiveness of React-OT on learning the TS through the transport perspective (Supplementary Figure \ref{Supp:guess_vs_ot_rmsd}).
Thanks to its deterministic character, React-OT generates a TS structure in one shot.
Consequently, the 0.39-second inference time corresponds to a 20-fold acceleration to single-shot OA-ReactDiff and near 1000-fold acceleration to 40-shot OA-ReactDiff with recommender\cite{OAReactDiff}, drastically accelerating the TS generation process at the same time as improving the precision.

\qquad Given that the initial guess of the sampling path deviates significantly from Gaussian distributions but remains relatively close to the TS, React-OT incorporates a rational implicit bias by assuming that the associations between initial guesses and true TSs are in proximity to the optimal coupling.
To further investigate the distance between React-OT and the true optimal transport paths, we gradually reduce the number of function evaluation (nfe) during the sampling process.
If the learned transition kernel resembles the true optimal transport, the path would be linear and thus a nfe=2 should suffice.
In practice, React-OT converges at nfe=6 for the RMSD of generated TS structures, which corresponds to an inference time of 0.05 seconds, demonstrating its effectiveness in learning the optimal transport path during structure generation\cite{I2SB} (Fig. \ref{fig:nfe}a).
However, we observe a much slower convergence at nfe=50 for the barrier height of generated TS structures(Supplementary Figure \ref{Supp:nfe_vs_time}).
This is likely due to the fact that the energy of TS, which usually contains strained bonds, are extremely sensitive to subtle structural changes in some directions compared to others.
Since this non-uniform response relationship between structure and energy is not captured by the scoring network, React-OT can hardly optimize the transport path for getting the exact energy, resulting a slower convergence in barrier height.
Consequently, React-OT with nfe=6 and nfe=200 generates structurally similar TSs, where most of their differences in RMSD is below a chemically meaningful threshold (that is, 0.01 Å, Fig. \ref{fig:nfe}b).
Yet the two sets of generated TSs exhibit a visible difference in their barrier height, with more than 80\% showing a deviation above 0.1 kcal/mol.
As both structural and energetic perspectives for generated TS are equally important, a nfe=50 is found sufficient when using React-OT in practice.

\qquad Despite that React-OT generates the TS structure in a fully deterministic manner, it is important to note that this is a desired behavior when both reactant and product conformations are fixed. For the same reaction with different reactant and product conformations, React-OT is able to generate TS structures that correspond to the input reaction conformation (Supplementary Figure \ref{Supp:reaction_conformations}). With the ability of generating accurate transition states and differentiating reaction conformations, React-OT can be effectively used to explore reaction networks from scratch. 
For instance, upon integrating React-OT with the Yet Another Reaction Program\cite{QiyuanNCS2021,zhao2022YARP2}—a reaction network exploration package which encompasses graph-based reaction enumeration rules and a comprehensive reaction conformational sampling algorithm\cite{zhao2022RCS}—we can explore the well-studied $\gamma$-ketohydroperoxide system, a common benchmark in recent research.\cite{grambow2018KHP,naz2020unimolecular,QiyuanNCS2021,zhao2022YARP2} The resulting two-step reaction network generated by React-OT was compared to previously published networks, demonstrating a high degree of similarity and successfully capturing all key reactions (Supplementary Section \ref{Supp:reaction_network}).

\begin{figure*}[t!]
    \centering
    \includegraphics[width=0.8\textwidth]{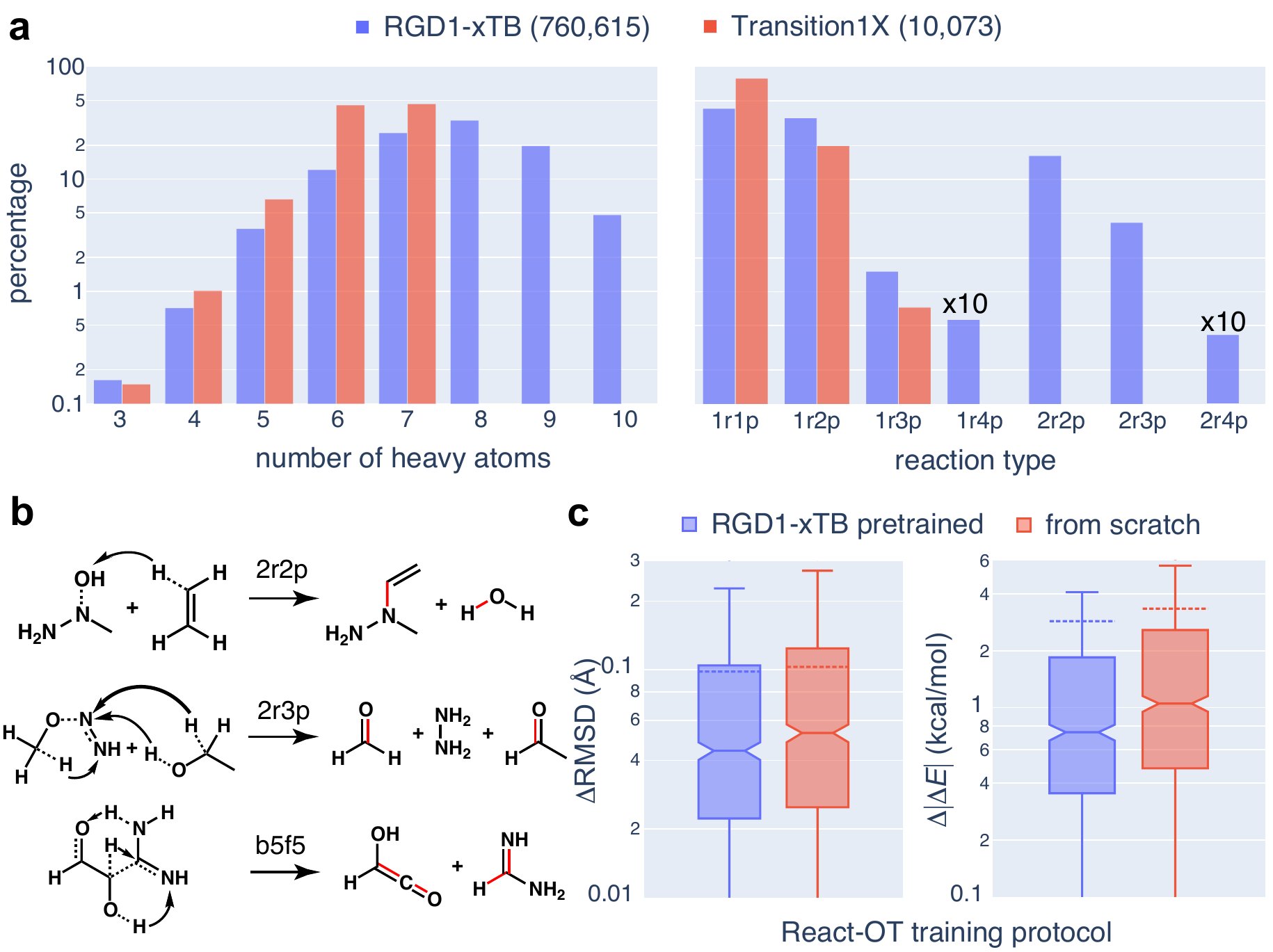}
    \caption{\textbf{React-OT with RGD1-xTB data for pretraining.}
    \textbf{a.} Distribution of number of heavy atoms (left) and reaction type (right) for Transition1X (red) and RGD1-xTB (blue). The percentages are shown in log scale for better visibility of rare cases. ArBp refers to a reaction that has A reactants and B products where a rule of A <= B is set. The percentage of 1r4p and 2r4p is multiplied by 10 to be shown on the plot.
    \textbf{b.} Examples of reaction types that are exclusively presented in RGD1-xTB.
    \textbf{c.} Box plot for RMSD (left) and $|\Delta \mathit{E}_\mathrm{TS}|$ (right) in log scale between the true and generated TS on 1,073 set-aside test reactions evaluated by React-OT trained from scratch on Transition1X (red) and React-OT with RGD1-xTB pretraining (blue). Both the mean (dashed horizontal line) and median (notch) are shown in the box plot. The nfe is 100 for this comparison.
    }
    \label{fig:pretraining}
\end{figure*}

\begin{table*}[th]
\centering 
\caption{\textbf{Summary of statistics for RMSD, barrier height error, and inference time of TSDiff, OA-ReactDiff and React-OT TS structures.} 
For React-OT, the inference time for nfe=6 and nfe=50 (in parenthesis) are both shown, corresponding to where the RMSD and barrier height error converge, respectively. 
For barrier height, besides mean and median, the percentage of reactions reaching the 1.58 kcal/mol chemical accuracy (\% chem. acc.) is also shown.
In case of OA-ReactDiff + recommender, OA-ReactDiff is run 40 times, where a final TS is selected by the recommender according to the ranking score. 
Note that TSDiff and NeuralNEB use different random seeds for partitioning train/test set compared to OA-ReactDiff and React-OT.
Note that the inference time of NeuralNEB was evaluated on CPU\cite{NeuralNEB}.
}
\resizebox{0.90\textwidth}{!}{
\begin{tabular}{l|cc|ccc|c}\toprule
\multicolumn{1}{c|}{Approach}&\multicolumn{2}{c|}{RMSD (Å)} &\multicolumn{3}{c|}{$|\Delta E_{\mathrm{TS}}|$ (kcal/mol)} &\multicolumn{1}{c}{Inference (sec)}\\\midrule
&mean &median &mean &median &\% chem. acc. \\\midrule
TSDiff&0.2526&0.2206&10.369&6.428&10.9&--\\
NeuralNEB&0.1358&0.0959&6.510&2.114&42.6&33.00\\
OA-ReactDiff&0.1800&0.0752&6.256&1.681&49.7&6.80\\
OA-ReactDiff + recommender&0.1297&0.0582&4.453&1.617&48.7&272.00\\
\hline \hline
React-OT&0.1029&0.0527&3.337&1.058&63.6&0.05 (0.39)\\
React-OT + pretraining&0.0981&0.0441&2.864&0.741&71.9&0.05 (0.39)\\
React-OT + pretraining + xTB RP&0.1056&0.0487&3.283&0.787&68.7&0.05 (0.39)\\
\bottomrule
\end{tabular}}
\label{table:summary}
\end{table*}

\vskip 0.1in
\paragraph{Pretraining on additional reactions further improves React-OT.}
Despite the availability of large computational datasets at DFT accuracy for molecular\cite{QM9} and materials\cite{MP} properties, surface-absorbate structures\cite{oc22_dataset}, and molecular dynamics trajectories\cite{MD17,MD22}, the size of datasets containing paired reactants, transition states (TSs), and products is often limited.\cite{ReactQM9,ts1x,RGD1}
This limitation is primarily due to the significantly higher computational cost—two orders of magnitude more expensive—associated with optimizing the minimum energy pathway for chemical reactions.\cite{RGD1} 
In contrast, reaction datasets at more affordable levels of theory, such as semi-empirical quantum chemistry, can be much larger. 
This disparity highlights the importance of developing efficient schemes for pretraining models with lower-accuracy reaction data.
For instance, during the generation of DFT-level reactions in RGD1, 760,615 intrinsic reaction coordination calculations were performed at the GFN2-xTB level, resulting in a dataset that can potentially serve as a valuable source of data (named as RGD1-xTB afterwards).\cite{RGD1} 
Compared to Transition1x, RGD1-xTB covers a much larger chemical space, with 57\% of reactions both consisting of more than seven heavy atoms (i.e., the largest system size in Transition1x) as well as more diverse bond rearrangement schemes during chemical reactions (Fig. \ref{fig:pretraining}a). 
Moreover, the dataset's reactants originate from PubChem, and generic enumeration rules were applied to construct the dataset, resulting in a more realistic and diverse set of chemical reactions involving multiple RP and breaking and forming of multiple bonds (Fig. \ref{fig:pretraining}b). 

\qquad With a dataset 75 times larger than Transition1x at hand, we pretrained React-OT on RGD1-xTB, which was then fine-tuned on the training set of Transition1x with a reduced learning rate. 
RGD1-xTB pretrained React-OT further improves on its quality of generated TS structures, yielding a 0.098Å mean RMSD and 2.86 kcal/mol mean barrier height error (Fig. \ref{fig:pretraining}c).
The mean error only improves by roughly 10\%, mostly due to the persistent high-error TSs generated by React-OT both with and without pretraining (Supplementary Figure \ref{Supp:persistent_high_error}). 
In contrast, the median of the structural RMSD and barrier height error is reduced by more than 25\%, suggesting the superior performance of React-OT after pretraining on RGD1-xTB (Table \ref{table:summary}). 
This pretraining scheme is helpful, in particular, in cases where the ground-truth reaction data (either DFT or experiment) is limited.
Considering that GFN2-xTB is roughly 3 orders of magnitude cheaper than DFT, this observation encourages the application of low-level theory on generating reaction datasets coupled with TS search algorithms (such as NEB), especially for large chemical systems where DFT is too computationally demanding.
In addition, we find great transferability of React-OT on out-of-distribution reactions, which gives similar performance on 41 Diels Alde reactions at different substrates and a standard test set of diverse reaction types that Birkholz et al. used to test TS optimization methods\cite{Birkholz-15}.

\begin{figure*}[t!]
    \centering
    \includegraphics[width=0.98\textwidth]{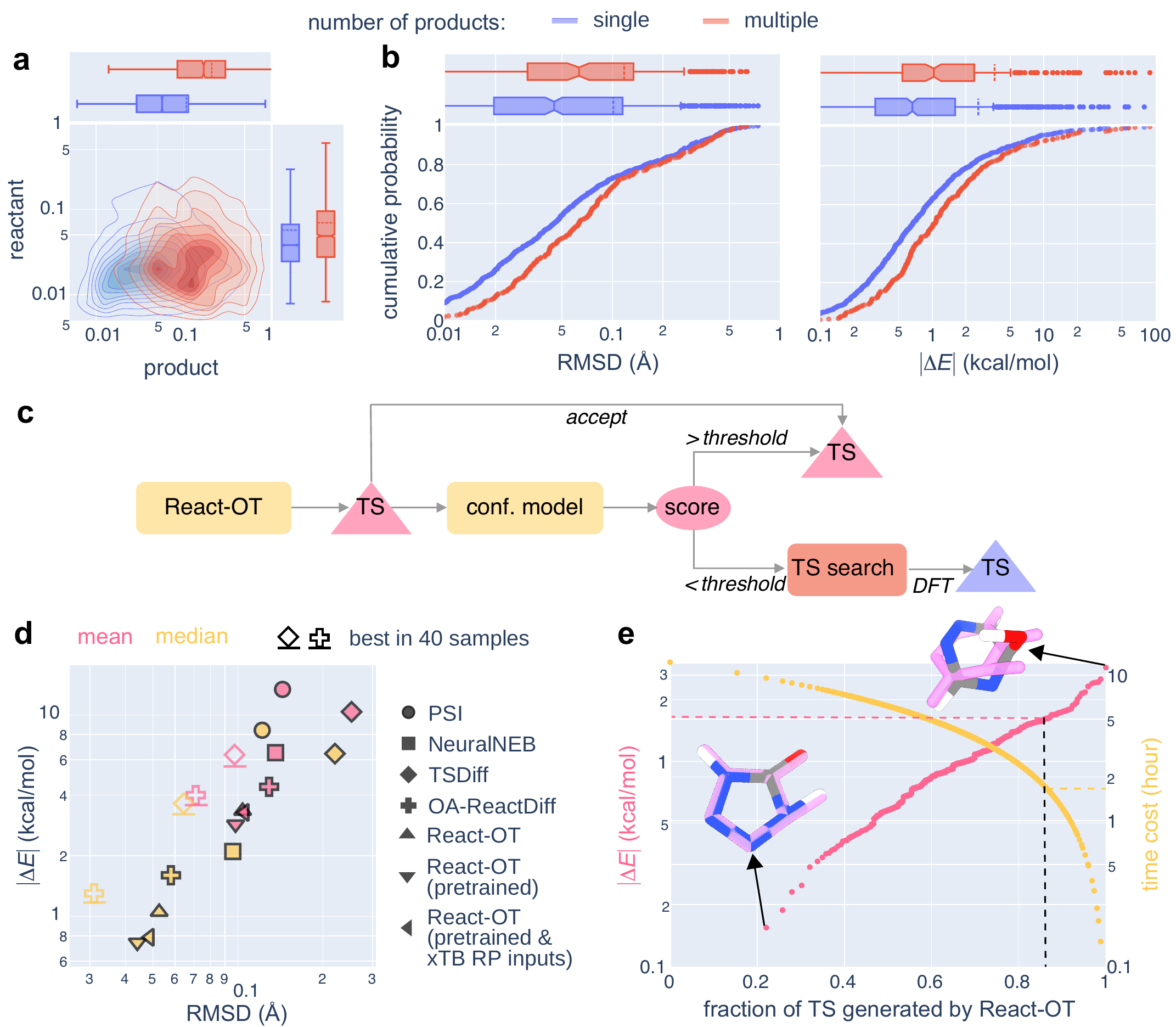}
    \caption{\textbf{React-OT with xTB RP as inputs and its performance in combination with DFT workflow and uncertainty control.}
    \textbf{a.} 2D contour plot for the RMSD of structures DFT- and xTB optimized RP. The single- (blue) and multi-product (red) reactions are shown separately.
    \textbf{b.} Cumulative probability for structure RMSD (left) and $|\Delta \mathit{E}_\mathrm{TS}|$ (right) between the true and React-OT generated TS on 1,073 set-aside test reactions. The single- (blue) and multi-product (red) reactions are shown separately. Both RMSD and $|\Delta \mathit{E}_\mathrm{TS}|$ are displaced in log scale for visibility of low error regime.
    \textbf{c.} Workflow of combining React-OT and conventional DFT-based TS search using a confidence (conf.) model. ML models are shown in orange squares, outputs from ML models are shown in pink triangles (TS) and pink circles (confidence score), DFT-based TS search shown in red square, with its output shown as blue triangle (true TS).
    \textbf{d.} RMSD versus $|\Delta \mathit{E}_\mathrm{TS}|$ for different TS generation approached where the mean absolute error is shown in pink and median absolute error is shown in yellow. OA-ReactDiff and TSDiff are also evaluated by the best sample among 40 sampling rounds (symbols without color filling), which, however, is not practical in real application settings.
    \textbf{e.} Average $|\Delta \mathit{E}_\mathrm{TS}|$ (pink, left y-axis) and time cost per reaction (yellow, right y-axis) with respect to the fraction of TS generated by React-OT under the control by uncertainty quantification. Dashed lines show the statistics at chemical accuracy (1.58 kcal/mol). Reference and generated TS structure (pink) for best and worst test reaction are shown. Atoms in the reference TS are colored as follows: C for gray; N for blue, O for red, and H for white.
    }
    \label{fig:xtbRP}
    
\end{figure*}

\vskip 0.1in
\paragraph{Utilizing xTB-optimized RP in React-OT.}
Usually the DFT-optimized RP are assumed known explicitly in double-ended TS search problem\cite{TSTReview}.
This assumption, however, is not practical in ML-accelerated workflow such as React-OT since the time spent on optimizing RP with DFT is several orders of magnitude longer than that for generating the TS structure.
Therefore, the robustness of being able to utilize approximate RP structures is of great value in ML-based TS search.
Here, we re-optimized RP in Transition1x with GFN2-xTB as an approximation to DFT-optimized structures.
These xTB-optimized RP still share some degrees of structural similarity to those optimized by DFT (that is, $\omega$B97x/6-31G(d)), especially for unimolecular reactions, with a mean RMSD of 0.07 Å (Fig. \ref{fig:xtbRP}a).
Reactions with multiple products, however, show a much larger RMSD (that is, 0.28 Å) for products optimized by GFN2-xTB and DFT (Supplementary Figure \ref{Supp:xtb_rmsd} and Supplementary Table \ref{Supp:table_xtb_rmsd}).
This large RMSD is mostly ascribed to the error in the alignment of multiple molecules in the product conformer.
Consequently, React-OT with xTB-optimized RP generates TS structures with both higher structural and energetic errors on reactions with multiple products than unimolecular reactions (Fig. \ref{fig:xtbRP}b).
However, despite using xTB-optimized RP that demonstrate a 0.07-0.28 Å mean RMSD as inputs, React-OT still generates TS structures at comparable performance, with a 0.106 Å mean RMSD and 0.049 Å median RMSD, only exhibiting less than 10\% increasing in RMSD compared to that using DFT-optimized RP as inputs (Table \ref{table:summary}).
In addition, React-OT still gives a RMSD as low as 0.12 Å for TSs at multi-molecular reactions, although there are large structural differences between xTB- and DFT-optimized products (mean RMSD of 0.28Å, Fig. \ref{fig:xtbRP}b).
The same behavior is observed on estimating barrier height, where React-OT also shows comparable performance when using xTB-optimized RP.
Despite a slight reduction in performance using xTB-optimized RP as inputs, React-OT still maintains state of the art compared to ML approaches with various flavors, such as diffusion models of 3D (OA-ReactDiff\cite{OAReactDiff}) and 2D RP (TSDiff\cite{2DTSDiff}), ML potential as surrogate in climbing-image NEB (NeuralNEB\cite{NeuralNEB}), and transformer-based generative models (PSI\cite{ChoiNatComm}) (Fig. \ref{fig:xtbRP}c, Supplementary Figure \ref{Supp:sumamry}).
This observation also suggests the optimal transport formalism used in React-OT still holds, at least approximately, on RP computed at different levels of theory.

\qquad There is often a need to integrate ML models in high throughput computational workflows to further accelerate data acquisition and chemical discovery\cite{KulikChemRev}.
With its great accuracy on generating TS structures even using xTB-optimized RP, we investigate the performance of React-OT integrated in typical high throughput computational workflows that mandate TS search.
Provided a pair of xTB-optimized RP, we first use React-OT to generate a TS structure (Fig. \ref{fig:xtbRP}e).
This TS structure is then passed through a confidence model to obtain a confidence score on the likelihood of reactants, React-OT predicted TS, and products forming an elementary reaction (see \textit{\nameref{training}}).
One would accept the React-OT structure if the confidence score is higher than a user-defined threshold, otherwise a conventional TS search calculation with DFT is performed.
With a confidence threshold of 0, all React-OT generated TS structures are preserved, giving a mean inference time of 0.39 second and barrier height error of 3.28 kcal/mol.
With a confidence threshold of 1, none of the React-OT generated TS structures is considered accurate, and thus one would run explicit TS search for all reactions, taking 12.8 hours per reaction on average.
By varying the confidence threshold, one can choose between being aggressive and conservative on activating React-OT in the computational workflow.
As we reduce the threshold, a larger fraction of TS structures are accepted from React-OT.
Meanwhile, the mean barrier height error increases smoothly and monotonically, which indicates the confidence model can successfully assess the quality of generated TS structures (Fig. \ref{fig:xtbRP}d, Supplementary Figure \ref{Supp:random_confidence}).
When the chemical accuracy of barrier height prediction (that is, 1.58 kcal/mol MAE) is targeted, 86\% TSs would be generated by React-OT and the remaining 14\% searched by climbing-image NEB with DFT, rendering a seven-fold acceleration compared to DFT-only computational workflows.
At the same time, an extremely low median structural RMSD of 0.031Å and median barrier height error of 0.54 kcal/mol would be achieved.
These observations showcase the promise of combining React-OT and conventional TS search algorithms in high throughput screening of elementary reactions and building large reaction networks.

\section*{Discussion}
Elucidating TS structures is essential for uncovering the underlying microscopic mechanisms of chemical reactions and estimating reaction barriers for building large reaction networks. 
We developed React-OT, an optimal transport approach for deterministically generating TS provided RP.
React-OT achieves great accuracy on generated TS structures within 0.4 seconds.
Pretrained on a large reaction dataset at GFN2-xTB level, React-OT further improves by 25\% in TS structural RMSD and barrier height.
In a practical setting where xTB-optimized RP are used as inputs, React-OT still shows comparable performance, generating TS at chemical accuracy with only one seventh of the cost for a hybrid ML-DFT high throughput computational workflow.

\qquad Despite the use of RP optimized by GFN2-xTB, the preparation of optimized RP is the most time consuming step (10 folds more expensive than others combined) in React-OT enabled TS search workflow. 
Therefore, ML approaches that accelerates RP conformer optimizations, either by an ML potential or with a 2D graph to 3D conformation sampling model, would be desired to unleash the potential of sub-second inference speed of React-OT in large reaction network exploration.
There, generic ML interatomic potentials (such as ANI-1xnr\cite{ANI-1xnr}, MACE-OFF23\cite{MACE-OFF23}, and DPA-2\cite{DPA-2}) would be helpful for a fast optimization for reactant and product structures.
In addition, there are ongoing efforts benchmarking\cite{M2Hub} the performance of these ML interatomic potentials systematically on TS optimization problem against diffusion and transport based generative modeling, which would be reported in a future work.
Due to the absence of large reaction datasets with charged species and metals, the demonstration of React-OT is limited to the scope of neutral organic chemical space with CNOH elements.
In future, it would be of great interest to generate more chemical reaction data and train React-OT on more diverse chemistry to enable its application on more realistic chemical systems, such as those with transition metal catalysts involved.

\qquad With the optimal transport approach, React-OT does not have any stochasticity in the sampling process, making it tailored for double-ended TS search.
To construct reaction networks, however, a model is required to generate products from reactants, which does not have a one-to-one mapping.
There, either diffusion models that carry some degree of stochasticity (for example, OA-ReactDiff\cite{OAReactDiff}) or conventional high throughput computational workflows that efficiently sample reactant-product pairs under xTB calculations (for example, YARP\cite{Zhao2021}), is preferred.
Although we only demonstrate React-OT on the TS search problem, there are many scenarios in biology, chemistry, and materials science, where an optimal transport setting is appropriate, including protein-ligand docking, molecule absorption on metal surfaces, and structural changes of materials in phase transitions.
For example, an extended transport approach based on Doob’s Lagrangian were recently developed find feasible transition paths on molecular simulation and protein folding tasks\cite{Doobs2024}.
We anticipate this optimal transport approach presented here insightful to applications in other domain of sciences.

\section*{Methods}

\paragraph{Equivariance in modeling chemical reactions}\label{equivariance}
A function $f$ is said to be equivariant to a group of actions $G$ if $g\circ f(x) = f(g\circ x)$ for any $g\in G$ acting on $x$~\cite{equivariance, gdlbook}.
In this paper, we specifically consider the Special Euclidean group in 3D space (SE(3)) which includes permutation, translation and rotation transformations.
We intentionally break the reflection symmetry so that our model can describe molecules with chirality.
In diffusion models, SE(3) equivariance is achieved by building an SE(3)-invariant prior and an SE(3)-equivariant transition kernel~\cite{frank}. However, in the dynamic optimal transport setting, we can design a more informative prior than Gaussian distribution~\cite{I2SB} (in this case, we use the linear interpolation between the reactant and product as an initial guess). As the paired data (initial guess and transition state) is available during training, we align the initial and target geometry for permutation, rotation, and reflection, and move the center of mass to the origin to remove the effect of translation. In addition, we have an equivariant transition kernel to transport from the initial to the target distribution.

\paragraph{Equivariant diffusion models}\label{diffusion}
Diffusion models are originally inspired from non-equilibrium thermodynamics~\cite{ddpm, diffusion2015,scoresde}.
A diffusion model has two processes, the forward (diffusing) process and the reverse (denoising) process.
The noise process gradually adds noise into the data until it becomes a prior (Gaussian) distribution:
$$
q(x_t|x_{t-1}) = \mathcal{N}(x_t|\alpha_t x_{t-1}, \sigma_t^2 I),
$$
where $\alpha_t$ controls the signal retained and $\sigma_t$ controls the noise added.
A signal-to-noise ratio is defined as $\text{SNR}(t)=\frac{\alpha_t^2}{\sigma_t^2}$. We set $\alpha_t = \sqrt{1-\sigma_t^2}$ following the variance preserving process in~\cite{scoresde}.

\qquad The \textit{true denoising process} can be written in a closed form due to the property of Gaussian noise:
\begin{align}
q(x_s|x_0, x_{t}) &= \mathcal{N}(x_s|\mu_{t\rightarrow s}(x_0, x_t), \sigma^2_{t\rightarrow s}I), \nonumber\\
\mu_{t\rightarrow s}(x_0, x_t) &= \frac{\alpha_{t|s}\sigma^2_s}{\sigma^2_t}x_t + \frac{\alpha_s\sigma^2_{t|s}}{\sigma^2_t}x \;\; \text{and} \;\; \sigma_{t\rightarrow s} = \frac{\sigma_{t|s}\sigma_{s}}{\sigma_t}, \nonumber
\end{align}
where
\textcolor{black}{
$s$ < $t$ refer to two different timesteps along the diffusion/denoising process ranging from 0 to T,
}
$\alpha_{t|s} = \frac{\alpha_{t}}{\alpha_{s}}$, $\sigma^2_{t|s} = \sigma^2_t - \alpha^2_{t|s} \sigma^2_{s}$.
However, this \textit{true denoising process} is dependent on $x_0$ which is the data distribution and not accessible.
Therefore, diffusion learns the denoising process by replacing $x_0$ with $x_t + \sigma_t \epsilon_{\theta}(x_t, t)$ predicted by a denoising network $\epsilon_{\theta}$, which predicts the difference of $x$ between two time steps (that is, $\epsilon$).
The training objective is to maximize the variational lower bound (VLB) on the likelihood of the training data:
$$
-\log p(x) \leq D_{KL}(q(x_T|x_0)||p_{\theta}(x_T)) - \log p(x_0|x_1) + \Sigma_{t=2}^T D_{KL}(q(x_{t-1}|x_0, x_t)|| p_{\theta}(x_{t-1}|x_t))
$$
Empirically, a simplified objective has been found to be efficient to optimize~\cite{ddpm}:
$$
\mathcal{L}_\mathrm{simple} = \frac{1}{2}||\epsilon - \epsilon_{\theta}(x_t, t)||^2,
$$
However, it is worth noting that diffusion model follows a protocol that transforms a fixed normal distribution to a target distribution. Therefore, one end of diffusion model can only be the normal distribution, which limits the use of prior knowledge in formulating the TS problem, motivating us to pursue flow matching objective and optimal transport formalism.

\paragraph{Details about React-OT}\label{ot}
\textit{Optimal transport.-- }
Optimal transport (OT) problems \cite{villani2009optimal,villani2021topics,santambrogio2015optimal,Linfeng2018MPFlow} investigate the most efficient means of transporting samples between two specified distributions.
Originating from Gaspard Monge's work \cite{monge1781memoire} in the 18th century, the problem has evolved and gained broader significance through Leonid Kantorovich's relaxation \cite{kantorovich1942translocation}.
Given a pair of boundary distributions $\mu,\nu \in \calP(\R^d)$, where $\calP(\R^d)$ denotes the space of probability distributions, the problem involves minimizing the, e.g., squared Euclidean, transport cost:
$$
W_2^2(\mu, \nu) \coloneqq \inf_{\pi \in \Pi(\mu,\nu)} \int\int \|x-y\|^2 \pi(x, y) \rd x \rd y,
$$
where $\Pi(\mu, \nu) \coloneqq \{ \pi \in \calP(\calX \times \calY): \int \pi(x,y)\rd y = \mu, \int \pi(x,y)\rd x = \nu \}$ is the set of couplings on $\R^d \times \R^d$ with respective marginals $\mu, \nu$.
The resulting distance $W_2^2(\mu, \nu)$ between $\mu$ and $\nu$, given the optimal transport coupling $\pi^\star$, is commonly referred to as the squared Wasserstein-2 distance.

\textit{Dynamic optimal transport.-- }
Dynamic optimal transport, as introduced by Benamou and Brenier~\cite{benamou2000computational}, presents the squared Wasserstein-2 distance $W_2^2(\mu, \nu)$ in a dynamic formulation.
This formulation involves optimizing the squared $L^2$ norm of a time-varying vector field $u_t(x_t)$, which transports samples from $x_0 \sim \mu(x_0)$ to $x_1 \sim \nu(x_1)$ according to an ODE, $\frac{dx_t}{dt} = u_t(x_t)$. Mathematically, this dynamic formulation is expressed as:
\begin{align}
W_2^2(\mu, \nu) \coloneqq& \inf_{q_t, u_t} \int_0^1\int \frac{1}{2}  \| u_t(x_t) \|^2 q_t(x_t) \rd x_t \rd t \nonumber \\
\text{s.t. } \fracpartial{q_t}{t}~+&~\nabla \cdot (u_t q_t) = 0, \quad q_0 = \mu, \quad q_1 = \nu \nonumber
\end{align}
subject to the continuity equation and the boundary conditions $q_0 = \mu$ and $q_1 = \nu$.
This approach describes a fully deterministic transportation process, distinguishing it from previous methods such as OA-ReactDiff, which rely on stochastic processes.
The minimal assumptions placed on the distributional boundaries $\mu, \nu$ in OT problems, in contrast to the Gaussian priors in standard diffusion models~\cite{ddpm}, enable the straightforward incorporation of domain-specific structures, such as the incorporation of pairing information, if available.

Given the OT coupling $\pi^\star(x, y)$, the optimal time-marginal $q_t^\star$ can be defined using the OT \emph{interpolant} \cite[Eq.~(7.8)]{peyre2017computational} as
$$
q_t^\star = P_{t\sharp} \pi^\star, \text{ where } P_t : (x,y) \mapsto (1-t) x + t y
$$
is the mapping of OT interpolant and $P_{t\sharp}$ is its corresponding push-forward operator.
In essence, the optimal trajectories between samples drawn from $(x_0, x_1) \sim \pi^\star$ move along straight lines $x_t = (1-t) x_0 + t x_1$.
Consequently, the optimal vector field can be constructed as a constant velocity $u_t^\star(x_t) = x_1 - x_0$.

\textit{Formulation of TS finding as optimal transport.-- }
Recent advances in flow matching methods \cite{FlowMatching,liu2022flow} have introduced an efficient framework for solving dynamic OT problems in high-dimensional Euclidean spaces. 
To harness this computational efficiency for molecular transportation, we reformulate the TS finding problem in $\R^d$, where the collections of ground-truth TS structures and their initial guesses are treated as $\nu \in \calP(\R^d)$ and $\mu \in \calP(\R^d)$, respectively. 
The dimensionality $d$ is determined by padding the maximum number of atoms in the collection. 
For instance, a collection with three carbon, two oxygen, and one hydrogen atom corresponds to an 18-dimensional space, as each atom occupies a 3D position. 
The state of each TS structure is represented by filling in atom coordinates according to their types, with unfilled coordinates set to a default large value. 
This formulation yields a well-posed OT problem in Euclidean space with an $L^2$ norm, where distances between molecules $x_0 \sim \mu$ and $x_1 \sim \nu$ are properly defined. Molecules with different atom compositions incur large distance values, making them suboptimal pairings. Unlike prior approaches for similar data structures, such as point clouds \cite{shen2021accurate} and graphs \cite{titouan2019optimal}, which often involve complex extensions to the OT framework, our formulation retains the scalability of the flow matching approach.

\textit{Practical implementation of flow matching.-- }

In practice, given a tuple of product (P), reactor (R), and transition state (TS), we set the initial and terminal states respectively to $x_0 = \frac{P + R}{2}$ $\sim \mu$ and $x_1 = TS$ $\sim \nu$. 
This is due to an implicit bias where the pairing information is assumed to be close to the optimal coupling, i.e., $(\frac{P+R}{2},TS) \sim \pi^\star$. This assumption generally holds when the pairings are sufficiently close to each other without much overlap with others~\cite{I2SB,somnath2023aligned}, which is indeed the case in the application being considered. 
Note that as R and P may differ by an arbitrary orientation, we always align them with an SE(3) group action by minimizing their RMSD using the Kabsch algorithm.
Given the pair of pre-aligned $(x_0, x_1)$ from $\pi^\star$, we leverage the construction of OT interpolant \cite[Eq.~(7.8)]{peyre2017computational} and minimize a {flow matching}~\cite{FlowMatching} objective:
$$
\min_{\theta} { \norm{u_\theta(x_t,t,z) - (x_1 - x_0) }^2},
$$
where $z$ encapsulates all conditional information required for predicting the flow matching target, such as the conformer structures of R and P.
Ablation study is performed to investigate the source of superior performance of React-OT, leading to results that both a good initial guess TS structure and optimal transport objective are important, leaving either out would result in a reduced performance (Supplementary Table \ref{Supp:table_ablation_study}).
At inference, we can sample $x_1 \sim \nu(x_1|x_0)$ by solving the parametrized ODE $\fracdiff{x_t}{t}= u^\star_{\theta}(x_t, t, z)$ using off-the-shelf numerical solvers.

\paragraph{Inpainting for conditional generation.}\label{inpainting}
Inpainting is a flexible technique to formulate the conditional generation problem for diffusion models. \cite{Repaint}
Instead of modeling the conditional distribution, inpainting models the joint distribution during training.
During inference, inpainting methods combine the conditional input as part of the context through the noising process of the diffusion model before denoising both the conditional input and the inpainting region together. 
The resampling technique~\cite{Repaint} has demonstrated excellent empirical performance in harmonizing the context of the denoising process as there is sometimes mismatch between the noised conditional input and the denoised inpainting region. 
However, at a given total step $T$, resampling increases the total number of sampling steps in each denoising step by sampling the inpainting region back and forth together with the conditional input.
For example, with a resample size of 10, and total step of 100, the number of function evaluation is 1,000 for a single sampling process, which is significantly larger than that in React-OT.

\paragraph{LEFTNet.}\label{leftnet}
We build our equivariant transition kernel on top of a recently proposed SE(3)-equivariant GNN, LEFTNet~\cite{leftnet}. LEFTNet achieves SE(3)-equivariance based on building local node- and edge-wise equivariant frames that scalarize vector (for example position, velocity) and higher order tensor (for example stress) geometric quantities. The geometric quantities are transformed back from scalars through a tensorization technique without loss of any information. LEFTNet is designed to handle Euclidean group symmetries including rotation, translation and reflection, as well as the permutation symmetry. To tailor the model for chemical reaction, we adopt the object-aware improvement from our previous work, OA-ReactDiff\cite{OAReactDiff}. In addition to handling symmetries, LEFTNet has strong geometric and function approximation expressiveness. Specifically, LEFTNet has a local structure encoding module which is proven to distinguish a hierarchy of local 3D isomorphism and a frame transition encoding module which is capable of learning universal equivariant functions. For more detailed descriptions, we refer reading the original LEFTNet\cite{leftnet} or OA-ReactDiff\cite{OAReactDiff} work.
It is noted, however, newly developed model architectures (for example, MACE\cite{MACE-MP0}) can be incorporated as the scoring network in React-OT as long as they fulfill the object-aware SE(3) symmetry.

\paragraph{Details for model training.}\label{training}

\textit{React-OT training.-- } We directly fine tuned the OA-ReactDiff model checkpoint with the new optimal transport objective and a reduced learning rate of 0.0001 (see \textit{\nameref{ot}}).
The scoring network, LEFTNet, has 96 radial basis functions, 196 hidden channels for message passing, 6 equivariant update blocks, and an interaction cuotff of 10Å.
The React-OT model was trained for an additional 200 epochs.

\textit{Confidence model training.-- } The confidence model is also fine tuned from the OA-ReactDiff model checkpoint, with the change of the final output layer to a\textit{sigmoid} function for predicting a probability ranging from 0 to 1 as the confidence score.
The OA-ReactDiff were run for 40 rounds of sampling on the 9,000 training reactions, which generated 360,000 synthetic reactions.
A reaction was labeled as "good" (that is, 1) if the generated TS structure has a RMSD < 0.2 Å compared to the true TS, otherwise was labeled as "bad" (that is, 0).
Here, we directly adopted the confidence model trained in OA-ReactDiff without further changes.
\section*{Code and data availability}
Code and data are currently under review and will be available as a open source repository on github.

\section*{Author contributions}
C.D., G.-H.L., and Y.D.: conceptualization, methodology, software, validation, investigation, data curation, writing of original draft, review and editing, and visualization. 
T.C.: methodology, software, and review and editing.
Q.Z.: dataset, writing of original draft, and review and editing.
H.J.: data curation, review and editing. 
C.P.G., E.A.T., and H.J.K.: writing of original draft, review and editing.

\section*{Acknowledgement}
G.H.L., T.C, and E.A.T. are supported by ARO Award \#W911NF2010151 and DoD Basic Research Office Award
HQ00342110002.
Y.D. and C.P.G. acknowledge funds from the Eric and Wendy Schmidt AI in Science Postdoctoral Fellowship, a Schmidt Futures program, the National Science Foundation (NSF), the Air Force Office of Scientific Research (AFOSR), the Toyota Research Institute (TRI), and the National Institute of Food and Agriculture (USDA/NIFA). 
H.J. and H.J.K. acknowledge funds from the  National Science Foundation (CBET-1846426) and the Office of Naval Research (N00014-20-1-2150).
C.D. and H.J. acknowledge the support for computational resources from MIT sandbox and AWS Activate program.

\section*{Competing interests}
The authors declare no competing financial interest at this moment.

\bibliographystyle{naturemag_doi}
\bibliography{main.bib}

\begin{thebibliography}{10}
\urlstyle{rm}
\expandafter\ifx\csname url\endcsname\relax
  \def\url#1{\texttt{#1}}\fi
\expandafter\ifx\csname urlprefix\endcsname\relax\def\urlprefix{URL }\fi
\expandafter\ifx\csname doiprefix\endcsname\relax\def\doiprefix{DOI: }\fi
\providecommand{\bibinfo}[2]{#2}
\providecommand{\eprint}[2][]{\url{#2}}

\bibitem{TSTReview}
\bibinfo{author}{Truhlar, D.~G.}, \bibinfo{author}{Garrett, B.~C.} \& \bibinfo{author}{Klippenstein, S.~J.}
\newblock \bibinfo{journal}{\bibinfo{title}{Current status of transition-state theory}}.
\newblock {\emph{\JournalTitle{Journal of Physical Chemistry}}} \textbf{\bibinfo{volume}{100}}, \bibinfo{pages}{12771--12800}, \doiprefix\url{10.1021/jp953748q} (\bibinfo{year}{1996}).
\newblock \eprint{https://doi.org/10.1021/jp953748q}.

\bibitem{EEricRev}
\bibinfo{author}{E, W.} \& \bibinfo{author}{Vanden-Eijnden, E.}
\newblock \bibinfo{journal}{\bibinfo{title}{Transition-path theory and path-finding algorithms for the study of rare events}}.
\newblock {\emph{\JournalTitle{Annual review of physical chemistry}}} \textbf{\bibinfo{volume}{61}}, \bibinfo{pages}{391—420}, \doiprefix\url{10.1146/annurev.physchem.040808.090412} (\bibinfo{year}{2010}).

\bibitem{ZimmermanWIREsRev}
\bibinfo{author}{Dewyer, A.~L.}, \bibinfo{author}{Argüelles, A.~J.} \& \bibinfo{author}{Zimmerman, P.~M.}
\newblock \bibinfo{journal}{\bibinfo{title}{Methods for exploring reaction space in molecular systems}}.
\newblock {\emph{\JournalTitle{WIREs Computational Molecular Science}}} \textbf{\bibinfo{volume}{8}}, \bibinfo{pages}{e1354}, \doiprefix\url{https://doi.org/10.1002/wcms.1354} (\bibinfo{year}{2018}).
\newblock \eprint{https://wires.onlinelibrary.wiley.com/doi/pdf/10.1002/wcms.1354}.

\bibitem{ReactNetworkAnnualRev}
\bibinfo{author}{Unsleber, J.~P.} \& \bibinfo{author}{Reiher, M.}
\newblock \bibinfo{journal}{\bibinfo{title}{The exploration of chemical reaction networks}}.
\newblock {\emph{\JournalTitle{Annual Review of Physical Chemistry}}} \textbf{\bibinfo{volume}{71}}, \bibinfo{pages}{121--142}, \doiprefix\url{10.1146/annurev-physchem-071119-040123} (\bibinfo{year}{2020}).
\newblock \bibinfo{note}{PMID: 32105566}, \eprint{https://doi.org/10.1146/annurev-physchem-071119-040123}.

\bibitem{Grzybowski2024}
\bibinfo{author}{Klucznik, T.} \emph{et~al.}
\newblock \bibinfo{journal}{\bibinfo{title}{Computational prediction of complex cationic rearrangement outcomes}}.
\newblock {\emph{\JournalTitle{Nature}}} \textbf{\bibinfo{volume}{625}}, \bibinfo{pages}{508--515}, \doiprefix\url{10.1038/s41586-023-06854-3} (\bibinfo{year}{2024}).

\bibitem{WalshDD2024}
\bibinfo{author}{Back, S.} \emph{et~al.}
\newblock \bibinfo{journal}{\bibinfo{title}{Accelerated chemical science with ai}}.
\newblock {\emph{\JournalTitle{Digital Discovery}}} \textbf{\bibinfo{volume}{3}}, \bibinfo{pages}{23--33}, \doiprefix\url{10.1039/D3DD00213F} (\bibinfo{year}{2024}).

\bibitem{KulikChemRev}
\bibinfo{author}{Nandy, A.} \emph{et~al.}
\newblock \bibinfo{journal}{\bibinfo{title}{Computational discovery of transition-metal complexes: From high-throughput screening to machine learning}}.
\newblock {\emph{\JournalTitle{Chemical Reviews}}} \textbf{\bibinfo{volume}{121}}, \bibinfo{pages}{9927--10000}, \doiprefix\url{10.1021/acs.chemrev.1c00347} (\bibinfo{year}{2021}).

\bibitem{NvJustinNatChem}
\bibinfo{author}{Zhang, S.} \emph{et~al.}
\newblock \bibinfo{journal}{\bibinfo{title}{Exploring the frontiers of condensed-phase chemistry with a general reactive machine learning potential}}.
\newblock {\emph{\JournalTitle{Nature Chemistry}}} \doiprefix\url{10.1038/s41557-023-01427-3} (\bibinfo{year}{2024}).

\bibitem{RWField2020}
\bibinfo{author}{Prozument, K.} \emph{et~al.}
\newblock \bibinfo{journal}{\bibinfo{title}{Photodissociation transition states characterized by chirped pulse millimeter wave spectroscopy}}.
\newblock {\emph{\JournalTitle{Proceedings of the National Academy of Sciences}}} \textbf{\bibinfo{volume}{117}}, \bibinfo{pages}{146--151}, \doiprefix\url{10.1073/pnas.1911326116} (\bibinfo{year}{2020}).
\newblock \eprint{https://www.pnas.org/doi/pdf/10.1073/pnas.1911326116}.

\bibitem{Wolf2023}
\bibinfo{author}{Liu, Y.} \emph{et~al.}
\newblock \bibinfo{journal}{\bibinfo{title}{Rehybridization dynamics into the pericyclic minimum of an electrocyclic reaction imaged in real-time}}.
\newblock {\emph{\JournalTitle{Nature Communications}}} \textbf{\bibinfo{volume}{14}}, \bibinfo{pages}{2795}, \doiprefix\url{10.1038/s41467-023-38513-6} (\bibinfo{year}{2023}).

\bibitem{DFTReview}
\bibinfo{author}{Mardirossian, N.} \& \bibinfo{author}{Head-Gordon, M.}
\newblock \bibinfo{journal}{\bibinfo{title}{Thirty years of density functional theory in computational chemistry: an overview and extensive assessment of 200 density functionals}}.
\newblock {\emph{\JournalTitle{Molecular Physics}}} \textbf{\bibinfo{volume}{115}}, \bibinfo{pages}{2315--2372}, \doiprefix\url{10.1080/00268976.2017.1333644} (\bibinfo{year}{2017}).
\newblock \eprint{https://doi.org/10.1080/00268976.2017.1333644}.

\bibitem{string_method}
\bibinfo{author}{Weinan, E.}, \bibinfo{author}{Ren, W.} \& \bibinfo{author}{Vanden-Eijnden, E.}
\newblock \bibinfo{journal}{\bibinfo{title}{String method for the study of rare events}}.
\newblock {\emph{\JournalTitle{Physical Review B}}} \textbf{\bibinfo{volume}{66}}, \bibinfo{pages}{052301} (\bibinfo{year}{2002}).

\bibitem{growing_string}
\bibinfo{author}{Peters, B.}, \bibinfo{author}{Heyden, A.}, \bibinfo{author}{Bell, A.~T.} \& \bibinfo{author}{Chakraborty, A.}
\newblock \bibinfo{journal}{\bibinfo{title}{A growing string method for determining transition states: Comparison to the nudged elastic band and string methods}}.
\newblock {\emph{\JournalTitle{The Journal of chemical physics}}} \textbf{\bibinfo{volume}{120}}, \bibinfo{pages}{7877--7886} (\bibinfo{year}{2004}).

\bibitem{NEB}
\bibinfo{author}{Sheppard, D.}, \bibinfo{author}{Terrell, R.} \& \bibinfo{author}{Henkelman, G.}
\newblock \bibinfo{journal}{\bibinfo{title}{Optimization methods for finding minimum energy paths}}.
\newblock {\emph{\JournalTitle{Journal of Chemical Physics}}} \textbf{\bibinfo{volume}{128}}, \bibinfo{pages}{134106}, \doiprefix\url{10.1063/1.2841941} (\bibinfo{year}{2008}).
\newblock \eprint{https://doi.org/10.1063/1.2841941}.

\bibitem{maeda2014AFIR}
\bibinfo{author}{Maeda, S.}, \bibinfo{author}{Taketsugu, T.} \& \bibinfo{author}{Morokuma, K.}
\newblock \bibinfo{journal}{\bibinfo{title}{Exploring transition state structures for intramolecular pathways by the artificial force induced reaction method}}.
\newblock {\emph{\JournalTitle{J. Comput. Chem.}}} \textbf{\bibinfo{volume}{35}}, \bibinfo{pages}{166--173} (\bibinfo{year}{2014}).

\bibitem{shang2013SSWM}
\bibinfo{author}{Shang, C.} \& \bibinfo{author}{Liu, Z.~P.}
\newblock \bibinfo{journal}{\bibinfo{title}{Stochastic surface walking method for structure prediction and pathway searching}}.
\newblock {\emph{\JournalTitle{J. Chem. Theory Comput.}}} \textbf{\bibinfo{volume}{9}}, \bibinfo{pages}{1838--1845} (\bibinfo{year}{2013}).

\bibitem{Durant1996}
\bibinfo{author}{Durant, J.~L.}
\newblock \bibinfo{journal}{\bibinfo{title}{Evaluation of transition state properties by density functional theory}}.
\newblock {\emph{\JournalTitle{Chemical Physics Letters}}} \textbf{\bibinfo{volume}{256}}, \bibinfo{pages}{595--602}, \doiprefix\url{0.1016/0009-2614(96)00478-2} (\bibinfo{year}{1996}).

\bibitem{zimmerman2013ZStucture}
\bibinfo{author}{Zimmerman, P.~M.}
\newblock \bibinfo{journal}{\bibinfo{title}{Automated discovery of chemically reasonable elementary reaction steps}}.
\newblock {\emph{\JournalTitle{J. Comput. Chem.}}} \textbf{\bibinfo{volume}{34}}, \bibinfo{pages}{1385--1392} (\bibinfo{year}{2013}).

\bibitem{Reiher2018}
\bibinfo{author}{Simm, G.~N.}, \bibinfo{author}{Vaucher, A.~C.} \& \bibinfo{author}{Reiher, M.}
\newblock \bibinfo{journal}{\bibinfo{title}{Exploration of reaction pathways and chemical transformation networks}}.
\newblock {\emph{\JournalTitle{Journal of Physical Chemistry A}}} \textbf{\bibinfo{volume}{123}}, \bibinfo{pages}{385--399}, \doiprefix\url{10.1021/acs.jpca.8b10007} (\bibinfo{year}{2019}).
\newblock \eprint{https://doi.org/10.1021/acs.jpca.8b10007}.

\bibitem{ReiherMRNetwork}
\bibinfo{author}{Unsleber, J.~P.} \emph{et~al.}
\newblock \bibinfo{journal}{\bibinfo{title}{High-throughput ab initio reaction mechanism exploration in the cloud with automated multi-reference validation}}.
\newblock {\emph{\JournalTitle{Journal of Physical Chemistry}}} \textbf{\bibinfo{volume}{158}}, \bibinfo{pages}{084803}, \doiprefix\url{10.1063/5.0136526} (\bibinfo{year}{2023}).
\newblock \eprint{https://doi.org/10.1063/5.0136526}.

\bibitem{QiyuanNCS2021}
\bibinfo{author}{Zhao, Q.} \& \bibinfo{author}{Savoie, B.~M.}
\newblock \bibinfo{journal}{\bibinfo{title}{Simultaneously improving reaction coverage and computational cost in automated reaction prediction tasks}}.
\newblock {\emph{\JournalTitle{Nature Computational Science}}} \textbf{\bibinfo{volume}{1}}, \bibinfo{pages}{479--490}, \doiprefix\url{10.1038/s43588-021-00101-3} (\bibinfo{year}{2021}).

\bibitem{Nanoreactor}
\bibinfo{author}{Wang, L.-P.} \emph{et~al.}
\newblock \bibinfo{journal}{\bibinfo{title}{Discovering chemistry with an ab initio nanoreactor}}.
\newblock {\emph{\JournalTitle{Nature Chemistry}}} \textbf{\bibinfo{volume}{6}}, \bibinfo{pages}{1044--1048}, \doiprefix\url{10.1038/nchem.2099} (\bibinfo{year}{2014}).

\bibitem{NA-Nanoreactor}
\bibinfo{author}{Pieri, E.} \emph{et~al.}
\newblock \bibinfo{journal}{\bibinfo{title}{The non-adiabatic nanoreactor: towards the automated discovery of photochemistry}}.
\newblock {\emph{\JournalTitle{Chem. Sci.}}} \textbf{\bibinfo{volume}{12}}, \bibinfo{pages}{7294--7307}, \doiprefix\url{10.1039/D1SC00775K} (\bibinfo{year}{2021}).

\bibitem{Zeng2020}
\bibinfo{author}{Zeng, J.}, \bibinfo{author}{Cao, L.}, \bibinfo{author}{Xu, M.}, \bibinfo{author}{Zhu, T.} \& \bibinfo{author}{Zhang, J. Z.~H.}
\newblock \bibinfo{journal}{\bibinfo{title}{Complex reaction processes in combustion unraveled by neural network-based molecular dynamics simulation}}.
\newblock {\emph{\JournalTitle{Nature Communications}}} \textbf{\bibinfo{volume}{11}}, \bibinfo{pages}{5713}, \doiprefix\url{10.1038/s41467-020-19497-z} (\bibinfo{year}{2020}).

\bibitem{KinBot}
\bibinfo{author}{{Van de Vijver}, R.} \& \bibinfo{author}{Zádor, J.}
\newblock \bibinfo{journal}{\bibinfo{title}{Kinbot: Automated stationary point search on potential energy surfaces}}.
\newblock {\emph{\JournalTitle{Computer Physics Communications}}} \textbf{\bibinfo{volume}{248}}, \bibinfo{pages}{106947}, \doiprefix\url{https://doi.org/10.1016/j.cpc.2019.106947} (\bibinfo{year}{2020}).

\bibitem{vonLilienfeld2020}
\bibinfo{author}{von Lilienfeld, O.~A.}, \bibinfo{author}{M{\"u}ller, K.-R.} \& \bibinfo{author}{Tkatchenko, A.}
\newblock \bibinfo{journal}{\bibinfo{title}{Exploring chemical compound space with quantum-based machine learning}}.
\newblock {\emph{\JournalTitle{Nature Reviews Chemistry}}} \textbf{\bibinfo{volume}{4}}, \bibinfo{pages}{347--358}, \doiprefix\url{10.1038/s41570-020-0189-9} (\bibinfo{year}{2020}).

\bibitem{Margraf2023}
\bibinfo{author}{Margraf, J.~T.}, \bibinfo{author}{Jung, H.}, \bibinfo{author}{Scheurer, C.} \& \bibinfo{author}{Reuter, K.}
\newblock \bibinfo{journal}{\bibinfo{title}{Exploring catalytic reaction networks with machine learning}}.
\newblock {\emph{\JournalTitle{Nature Catalysis}}} \textbf{\bibinfo{volume}{6}}, \bibinfo{pages}{112--121}, \doiprefix\url{10.1038/s41929-022-00896-y} (\bibinfo{year}{2023}).

\bibitem{NeuralNEB}
\bibinfo{author}{Schreiner, M.}, \bibinfo{author}{Bhowmik, A.}, \bibinfo{author}{Vegge, T.}, \bibinfo{author}{Jørgensen, P.~B.} \& \bibinfo{author}{Winther, O.}
\newblock \bibinfo{journal}{\bibinfo{title}{Neuralneb—neural networks can find reaction paths fast}}.
\newblock {\emph{\JournalTitle{Machine Learning: Science and Technology}}} \textbf{\bibinfo{volume}{3}}, \bibinfo{pages}{045022}, \doiprefix\url{10.1088/2632-2153/aca23e} (\bibinfo{year}{2022}).

\bibitem{ANI-1xnr}
\bibinfo{author}{Zhang, S.} \emph{et~al.}
\newblock \bibinfo{journal}{\bibinfo{title}{Exploring the frontiers of condensed-phase chemistry with a general reactive machine learning potential}}.
\newblock {\emph{\JournalTitle{Nature Chemistry}}} \textbf{\bibinfo{volume}{16}}, \bibinfo{pages}{727--734}, \doiprefix\url{10.1038/s41557-023-01427-3} (\bibinfo{year}{2024}).

\bibitem{RLTS}
\bibinfo{author}{Zhang, J.} \emph{et~al.}
\newblock \bibinfo{journal}{\bibinfo{title}{Deep reinforcement learning of transition states}}.
\newblock {\emph{\JournalTitle{Phys. Chem. Chem. Phys.}}} \textbf{\bibinfo{volume}{23}}, \bibinfo{pages}{6888--6895}, \doiprefix\url{10.1039/D0CP06184K} (\bibinfo{year}{2021}).

\bibitem{pips}
\bibinfo{author}{Holdijk, L.} \emph{et~al.}
\newblock \bibinfo{journal}{\bibinfo{title}{Stochastic optimal control for collective variable free sampling of molecular transition paths}}.
\newblock {\emph{\JournalTitle{Advances in Neural Information Processing Systems}}} \textbf{\bibinfo{volume}{36}} (\bibinfo{year}{2024}).

\bibitem{GreenPCCP}
\bibinfo{author}{Pattanaik, L.}, \bibinfo{author}{Ingraham, J.~B.}, \bibinfo{author}{Grambow, C.~A.} \& \bibinfo{author}{Green, W.~H.}
\newblock \bibinfo{journal}{\bibinfo{title}{Generating transition states of isomerization reactions with deep learning}}.
\newblock {\emph{\JournalTitle{Phys. Chem. Chem. Phys.}}} \textbf{\bibinfo{volume}{22}}, \bibinfo{pages}{23618--23626}, \doiprefix\url{10.1039/D0CP04670A} (\bibinfo{year}{2020}).

\bibitem{EquiReact}
\bibinfo{author}{van Gerwen, P.} \emph{et~al.}
\newblock \bibinfo{title}{Equireact: An equivariant neural network for chemical reactions} (\bibinfo{year}{2023}).
\newblock \eprint{2312.08307}.

\bibitem{TSGAN}
\bibinfo{author}{Makoś, M.~Z.}, \bibinfo{author}{Verma, N.}, \bibinfo{author}{Larson, E.~C.}, \bibinfo{author}{Freindorf, M.} \& \bibinfo{author}{Kraka, E.}
\newblock \bibinfo{journal}{\bibinfo{title}{Generative adversarial networks for transition state geometry prediction}}.
\newblock {\emph{\JournalTitle{Journal of Chemical Physics}}} \textbf{\bibinfo{volume}{155}}, \bibinfo{pages}{024116}, \doiprefix\url{10.1063/5.0055094} (\bibinfo{year}{2021}).
\newblock \eprint{https://doi.org/10.1063/5.0055094}.

\bibitem{ChoiNatComm}
\bibinfo{author}{Choi, S.}
\newblock \bibinfo{journal}{\bibinfo{title}{Prediction of transition state structures of gas-phase chemical reactions via machine learning}}.
\newblock {\emph{\JournalTitle{Nature Communications}}} \textbf{\bibinfo{volume}{14}}, \bibinfo{pages}{1168}, \doiprefix\url{10.1038/s41467-023-36823-3} (\bibinfo{year}{2023}).

\bibitem{ddpm}
\bibinfo{author}{Ho, J.}, \bibinfo{author}{Jain, A.} \& \bibinfo{author}{Abbeel, P.}
\newblock \bibinfo{title}{Denoising diffusion probabilistic models}.
\newblock In \bibinfo{editor}{Larochelle, H.}, \bibinfo{editor}{Ranzato, M.}, \bibinfo{editor}{Hadsell, R.}, \bibinfo{editor}{Balcan, M.} \& \bibinfo{editor}{Lin, H.} (eds.) \emph{\bibinfo{booktitle}{Advances in Neural Information Processing Systems}}, vol.~\bibinfo{volume}{33}, \bibinfo{pages}{6840--6851} (\bibinfo{publisher}{Curran Associates, Inc.}, \bibinfo{year}{2020}).

\bibitem{2DTSDiff}
\bibinfo{author}{Kim, S.}, \bibinfo{author}{Woo, J.} \& \bibinfo{author}{Kim, W.~Y.}
\newblock \bibinfo{title}{Diffusion-based generative ai for exploring transition states from 2d molecular graphs} (\bibinfo{year}{2023}).
\newblock \eprint{2304.12233}.

\bibitem{OAReactDiff}
\bibinfo{author}{Duan, C.}, \bibinfo{author}{Du, Y.}, \bibinfo{author}{Jia, H.} \& \bibinfo{author}{Kulik, H.~J.}
\newblock \bibinfo{journal}{\bibinfo{title}{Accurate transition state generation with an object-aware equivariant elementary reaction diffusion model}}.
\newblock {\emph{\JournalTitle{Nature Computational Science}}} \textbf{\bibinfo{volume}{3}}, \bibinfo{pages}{1045--1055}, \doiprefix\url{10.1038/s43588-023-00563-7} (\bibinfo{year}{2023}).

\bibitem{AlanDiffusion2024}
\bibinfo{author}{Cheng, A.~H.}, \bibinfo{author}{Lo, A.}, \bibinfo{author}{Miret, S.}, \bibinfo{author}{Pate, B.~H.} \& \bibinfo{author}{Aspuru-Guzik, A.}
\newblock \bibinfo{journal}{\bibinfo{title}{{Determining 3D structure from molecular formula and isotopologue rotational spectra in natural abundance with reflection-equivariant diffusion}}}.
\newblock {\emph{\JournalTitle{The Journal of Chemical Physics}}} \textbf{\bibinfo{volume}{160}}, \bibinfo{pages}{124115}, \doiprefix\url{10.1063/5.0196620} (\bibinfo{year}{2024}).
\newblock \eprint{https://pubs.aip.org/aip/jcp/article-pdf/doi/10.1063/5.0196620/19855901/124115\_1\_5.0196620.pdf}.

\bibitem{DFARec}
\bibinfo{author}{Duan, C.}, \bibinfo{author}{Nandy, A.}, \bibinfo{author}{Meyer, R.}, \bibinfo{author}{Arunachalam, N.} \& \bibinfo{author}{Kulik, H.~J.}
\newblock \bibinfo{journal}{\bibinfo{title}{A transferable recommender approach for selecting the best density functional approximations in chemical discovery}}.
\newblock {\emph{\JournalTitle{Nature Computational Science}}} \textbf{\bibinfo{volume}{3}}, \bibinfo{pages}{38--47}, \doiprefix\url{10.1038/s43588-022-00384-0} (\bibinfo{year}{2023}).

\bibitem{DiffDock}
\bibinfo{author}{Corso, G.}, \bibinfo{author}{Stärk, H.}, \bibinfo{author}{Jing, B.}, \bibinfo{author}{Barzilay, R.} \& \bibinfo{author}{Jaakkola, T.}
\newblock \bibinfo{journal}{\bibinfo{title}{Diff{D}ock: Diffusion steps, twists, and turns for molecular docking}}.
\newblock {\emph{\JournalTitle{arXiv:2210.01776}}}  (\bibinfo{year}{2023}).

\bibitem{RGD1}
\bibinfo{author}{Zhao, Q.} \emph{et~al.}
\newblock \bibinfo{journal}{\bibinfo{title}{Comprehensive exploration of graphically defined reaction spaces}}.
\newblock {\emph{\JournalTitle{Scientific Data}}} \textbf{\bibinfo{volume}{10}}, \bibinfo{pages}{145}, \doiprefix\url{10.1038/s41597-023-02043-z} (\bibinfo{year}{2023}).

\bibitem{GFN2-xTB}
\bibinfo{author}{Bannwarth, C.}, \bibinfo{author}{Ehlert, S.} \& \bibinfo{author}{Grimme, S.}
\newblock \bibinfo{journal}{\bibinfo{title}{Gfn2-xtb---an accurate and broadly parametrized self-consistent tight-binding quantum chemical method with multipole electrostatics and density-dependent dispersion contributions}}.
\newblock {\emph{\JournalTitle{Journal of Chemical Theory and Computation}}} \textbf{\bibinfo{volume}{15}}, \bibinfo{pages}{1652--1671}, \doiprefix\url{10.1021/acs.jctc.8b01176} (\bibinfo{year}{2019}).

\bibitem{diffusion2015}
\bibinfo{author}{Sohl-Dickstein, J.}, \bibinfo{author}{Weiss, E.}, \bibinfo{author}{Maheswaranathan, N.} \& \bibinfo{author}{Ganguli, S.}
\newblock \bibinfo{title}{Deep unsupervised learning using nonequilibrium thermodynamics}.
\newblock In \emph{\bibinfo{booktitle}{International Conference on Machine Learning}}, \bibinfo{pages}{2256--2265} (\bibinfo{year}{2015}).

\bibitem{scoresde}
\bibinfo{author}{Song, Y.} \emph{et~al.}
\newblock \bibinfo{title}{Score-based generative modeling through stochastic differential equations}.
\newblock In \emph{\bibinfo{booktitle}{International Conference on Learning Representations}} (\bibinfo{year}{2020}).

\bibitem{FlowMatching}
\bibinfo{author}{Lipman, Y.}, \bibinfo{author}{Chen, R. T.~Q.}, \bibinfo{author}{Ben-Hamu, H.}, \bibinfo{author}{Nickel, M.} \& \bibinfo{author}{Le, M.}
\newblock \bibinfo{title}{Flow matching for generative modeling}.
\newblock In \emph{\bibinfo{booktitle}{The Eleventh International Conference on Learning Representations}} (\bibinfo{year}{2023}).

\bibitem{I2SB}
\bibinfo{author}{Liu, G.-H.} \emph{et~al.}
\newblock \bibinfo{title}{{I{$^2$}SB: Image-to-Image Schr{\"o}dinger bridge}}.
\newblock In \emph{\bibinfo{booktitle}{International Conference on Machine Learning}} (\bibinfo{year}{2023}).

\bibitem{somnath2023aligned}
\bibinfo{author}{Somnath, V.~R.} \emph{et~al.}
\newblock \bibinfo{title}{Aligned diffusion schr$\backslash$" odinger bridges}.
\newblock In \emph{\bibinfo{booktitle}{Conference on Uncertainty in Artificial Intelligence}} (\bibinfo{year}{2023}).

\bibitem{zhao2022RCS}
\bibinfo{author}{Zhao, Q.}, \bibinfo{author}{Hsu, H.-H.} \& \bibinfo{author}{Savoie, B.}
\newblock \bibinfo{journal}{\bibinfo{title}{Conformational sampling for transition state searches on a computational budget}}.
\newblock {\emph{\JournalTitle{J. Chem. Theory Comput.}}} \textbf{\bibinfo{volume}{18}}, \bibinfo{pages}{3006--3016} (\bibinfo{year}{2022}).

\bibitem{sindhu2019theoretical}
\bibinfo{author}{Sindhu, A.}, \bibinfo{author}{Pradhan, R.}, \bibinfo{author}{Lourderaj, U.} \& \bibinfo{author}{Paranjothy, M.}
\newblock \bibinfo{journal}{\bibinfo{title}{Theoretical investigation of the isomerization pathways of diazenes: torsion vs. inversion}}.
\newblock {\emph{\JournalTitle{Phys. Chem. Chem. Phys.}}} \textbf{\bibinfo{volume}{21}}, \bibinfo{pages}{15678--15685} (\bibinfo{year}{2019}).

\bibitem{koda2024locating}
\bibinfo{author}{Koda, S.-i.} \& \bibinfo{author}{Saito, S.}
\newblock \bibinfo{journal}{\bibinfo{title}{Locating transition states by variational reaction path optimization with an energy-derivative-free objective function}}.
\newblock {\emph{\JournalTitle{J. Chem. Theory Comput.}}} \textbf{\bibinfo{volume}{20}}, \bibinfo{pages}{2798--2811} (\bibinfo{year}{2024}).

\bibitem{ts1x}
\bibinfo{author}{Schreiner, M.}, \bibinfo{author}{Bhowmik, A.}, \bibinfo{author}{Vegge, T.}, \bibinfo{author}{Busk, J.} \& \bibinfo{author}{Winther, O.}
\newblock \bibinfo{journal}{\bibinfo{title}{Transition1x - a dataset for building generalizable reactive machine learning potentials}}.
\newblock {\emph{\JournalTitle{Scientific Data}}} \textbf{\bibinfo{volume}{9}}, \bibinfo{pages}{779}, \doiprefix\url{10.1038/s41597-022-01870-w} (\bibinfo{year}{2022}).

\bibitem{CINEB}
\bibinfo{author}{Henkelman, G.}, \bibinfo{author}{Uberuaga, B.~P.} \& \bibinfo{author}{Jónsson, H.}
\newblock \bibinfo{journal}{\bibinfo{title}{A climbing image nudged elastic band method for finding saddle points and minimum energy paths}}.
\newblock {\emph{\JournalTitle{Journal of Physical Chemistry}}} \textbf{\bibinfo{volume}{113}}, \bibinfo{pages}{9901--9904}, \doiprefix\url{10.1063/1.1329672} (\bibinfo{year}{2000}).
\newblock \eprint{https://doi.org/10.1063/1.1329672}.

\bibitem{wb97x}
\bibinfo{author}{Chai, J.-D.} \& \bibinfo{author}{Head-Gordon, M.}
\newblock \bibinfo{journal}{\bibinfo{title}{Systematic optimization of long-range corrected hybrid density functionals}}.
\newblock {\emph{\JournalTitle{Journal of Physical Chemistry}}} \textbf{\bibinfo{volume}{128}}, \bibinfo{pages}{084106}, \doiprefix\url{10.1063/1.2834918} (\bibinfo{year}{2008}).
\newblock \eprint{https://doi.org/10.1063/1.2834918}.

\bibitem{631gs}
\bibinfo{author}{Ditchfield, R.}, \bibinfo{author}{Hehre, W.~J.} \& \bibinfo{author}{Pople, J.~A.}
\newblock \bibinfo{journal}{\bibinfo{title}{Self‐consistent molecular‐orbital methods. ix. an extended gaussian‐type basis for molecular‐orbital studies of organic molecules}}.
\newblock {\emph{\JournalTitle{Journal of Physical Chemistry}}} \textbf{\bibinfo{volume}{54}}, \bibinfo{pages}{724--728}, \doiprefix\url{10.1063/1.1674902} (\bibinfo{year}{1971}).
\newblock \eprint{https://doi.org/10.1063/1.1674902}.

\bibitem{Grambow2020}
\bibinfo{author}{Grambow, C.~A.}, \bibinfo{author}{Pattanaik, L.} \& \bibinfo{author}{Green, W.~H.}
\newblock \bibinfo{journal}{\bibinfo{title}{Reactants, products, and transition states of elementary chemical reactions based on quantum chemistry}}.
\newblock {\emph{\JournalTitle{Scientific Data}}} \textbf{\bibinfo{volume}{7}}, \bibinfo{pages}{137}, \doiprefix\url{10.1038/s41597-020-0460-4} (\bibinfo{year}{2020}).

\bibitem{GreenDataApp}
\bibinfo{author}{Grambow, C.~A.}, \bibinfo{author}{Pattanaik, L.} \& \bibinfo{author}{Green, W.~H.}
\newblock \bibinfo{journal}{\bibinfo{title}{Deep learning of activation energies}}.
\newblock {\emph{\JournalTitle{Journal of Physical Chemistry Letters}}} \textbf{\bibinfo{volume}{11}}, \bibinfo{pages}{2992--2997}, \doiprefix\url{10.1021/acs.jpclett.0c00500} (\bibinfo{year}{2020}).
\newblock \eprint{https://doi.org/10.1021/acs.jpclett.0c00500}.

\bibitem{gdb17}
\bibinfo{author}{Ruddigkeit, L.}, \bibinfo{author}{van Deursen, R.}, \bibinfo{author}{Blum, L.~C.} \& \bibinfo{author}{Reymond, J.-L.}
\newblock \bibinfo{journal}{\bibinfo{title}{Enumeration of 166 billion organic small molecules in the chemical universe database {GDB}-17}}.
\newblock {\emph{\JournalTitle{Journal of Chemical Information and Modeling}}} \textbf{\bibinfo{volume}{52}}, \bibinfo{pages}{2864--2875}, \doiprefix\url{10.1021/ci300415d} (\bibinfo{year}{2012}).
\newblock \eprint{https://doi.org/10.1021/ci300415d}.

\bibitem{leftnet}
\bibinfo{author}{Du, W.} \emph{et~al.}
\newblock \bibinfo{journal}{\bibinfo{title}{A new perspective on building efficient and expressive 3{D} equivariant graph neural networks}}.
\newblock {\emph{\JournalTitle{arXiv:2304.04757}}}  (\bibinfo{year}{2023}).

\bibitem{ChemAccFu2022}
\bibinfo{author}{Fu, H.}, \bibinfo{author}{Zhou, Y.}, \bibinfo{author}{Jing, X.}, \bibinfo{author}{Shao, X.} \& \bibinfo{author}{Cai, W.}
\newblock \bibinfo{journal}{\bibinfo{title}{Meta-analysis reveals that absolute binding free-energy calculations approach chemical accuracy}}.
\newblock {\emph{\JournalTitle{Journal of Medicinal Chemistry}}} \textbf{\bibinfo{volume}{65}}, \bibinfo{pages}{12970--12978}, \doiprefix\url{10.1021/acs.jmedchem.2c00796} (\bibinfo{year}{2022}).

\bibitem{ChemAccAdamo2022}
\bibinfo{author}{Bremond, E.}, \bibinfo{author}{Li, H.}, \bibinfo{author}{Perez-Jimenez, A.~J.}, \bibinfo{author}{Sancho-Garcia, J.~C.} \& \bibinfo{author}{Adamo, C.}
\newblock \bibinfo{journal}{\bibinfo{title}{{Tackling an accurate description of molecular reactivity with double-hybrid density functionals}}}.
\newblock {\emph{\JournalTitle{The Journal of Chemical Physics}}} \textbf{\bibinfo{volume}{156}}, \bibinfo{pages}{161101}, \doiprefix\url{10.1063/5.0087586} (\bibinfo{year}{2022}).
\newblock \eprint{https://pubs.aip.org/aip/jcp/article-pdf/doi/10.1063/5.0087586/16539886/161101\_1\_online.pdf}.

\bibitem{zhao2022YARP2}
\bibinfo{author}{Zhao, Q.} \& \bibinfo{author}{Savoie, B.~M.}
\newblock \bibinfo{journal}{\bibinfo{title}{Algorithmic explorations of unimolecular and bimolecular reaction spaces}}.
\newblock {\emph{\JournalTitle{Angew. Chem., Int. Ed.}}} \textbf{\bibinfo{volume}{61}}, \bibinfo{pages}{e202210693}, \doiprefix\url{https://doi.org/10.1002/anie.202210693} (\bibinfo{year}{2022}).

\bibitem{grambow2018KHP}
\bibinfo{author}{Grambow, C.~A.} \emph{et~al.}
\newblock \bibinfo{journal}{\bibinfo{title}{Unimolecular reaction pathways of a $\gamma$-ketohydroperoxide from combined application of automated reaction discovery methods}}.
\newblock {\emph{\JournalTitle{J. Am. Chem. Soc.}}} \textbf{\bibinfo{volume}{140}}, \bibinfo{pages}{1035--1048} (\bibinfo{year}{2018}).

\bibitem{naz2020unimolecular}
\bibinfo{author}{Naz, E.~G.} \& \bibinfo{author}{Paranjothy, M.}
\newblock \bibinfo{journal}{\bibinfo{title}{Unimolecular dissociation of $\gamma$-ketohydroperoxide via direct chemical dynamics simulations}}.
\newblock {\emph{\JournalTitle{J Phys. Chem. A}}} \textbf{\bibinfo{volume}{124}}, \bibinfo{pages}{8120--8127} (\bibinfo{year}{2020}).

\bibitem{QM9}
\bibinfo{author}{Ramakrishnan, R.}, \bibinfo{author}{Dral, P.~O.}, \bibinfo{author}{Rupp, M.} \& \bibinfo{author}{von Lilienfeld, O.~A.}
\newblock \bibinfo{journal}{\bibinfo{title}{Quantum chemistry structures and properties of 134 kilo molecules}}.
\newblock {\emph{\JournalTitle{Scientific Data}}} \textbf{\bibinfo{volume}{1}}, \bibinfo{pages}{140022}, \doiprefix\url{10.1038/sdata.2014.22} (\bibinfo{year}{2014}).

\bibitem{MP}
\bibinfo{author}{Jain, A.} \emph{et~al.}
\newblock \bibinfo{journal}{\bibinfo{title}{{Commentary: The Materials Project: A materials genome approach to accelerating materials innovation}}}.
\newblock {\emph{\JournalTitle{APL Materials}}} \textbf{\bibinfo{volume}{1}}, \bibinfo{pages}{011002}, \doiprefix\url{10.1063/1.4812323} (\bibinfo{year}{2013}).
\newblock \eprint{https://pubs.aip.org/aip/apm/article-pdf/doi/10.1063/1.4812323/13163869/011002\_1\_online.pdf}.

\bibitem{oc22_dataset}
\bibinfo{author}{Tran*, R.} \emph{et~al.}
\newblock \bibinfo{journal}{\bibinfo{title}{The open catalyst 2022 (oc22) dataset and challenges for oxide electrocatalysts}}.
\newblock {\emph{\JournalTitle{ACS Catalysis}}}  (\bibinfo{year}{2023}).

\bibitem{MD17}
\bibinfo{author}{Chmiela, S.} \emph{et~al.}
\newblock \bibinfo{journal}{\bibinfo{title}{Machine learning of accurate energy-conserving molecular force fields}}.
\newblock {\emph{\JournalTitle{Science Advances}}} \textbf{\bibinfo{volume}{3}}, \bibinfo{pages}{e1603015}, \doiprefix\url{10.1126/sciadv.1603015} (\bibinfo{year}{2017}).
\newblock \eprint{https://www.science.org/doi/pdf/10.1126/sciadv.1603015}.

\bibitem{MD22}
\bibinfo{author}{Chmiela, S.} \emph{et~al.}
\newblock \bibinfo{journal}{\bibinfo{title}{Accurate global machine learning force fields for molecules with hundreds of atoms}}.
\newblock {\emph{\JournalTitle{Science Advances}}} \textbf{\bibinfo{volume}{9}}, \bibinfo{pages}{eadf0873}, \doiprefix\url{10.1126/sciadv.adf0873} (\bibinfo{year}{2023}).
\newblock \eprint{https://www.science.org/doi/pdf/10.1126/sciadv.adf0873}.

\bibitem{ReactQM9}
\bibinfo{author}{Nandi, S.}, \bibinfo{author}{Vegge, T.} \& \bibinfo{author}{Bhowmik, A.}
\newblock \bibinfo{journal}{\bibinfo{title}{Multixc-qm9: Large dataset of molecular and reaction energies from multi-level quantum chemical methods}}.
\newblock {\emph{\JournalTitle{Scientific Data}}} \textbf{\bibinfo{volume}{10}}, \bibinfo{pages}{783}, \doiprefix\url{10.1038/s41597-023-02690-2} (\bibinfo{year}{2023}).

\bibitem{Birkholz-15}
\bibinfo{author}{Birkholz, A.~B.} \& \bibinfo{author}{Schlegel, H.~B.}
\newblock \bibinfo{journal}{\bibinfo{title}{Using bonding to guide transition state optimization}}.
\newblock {\emph{\JournalTitle{Journal of Computational Chemistry}}} \textbf{\bibinfo{volume}{36}}, \bibinfo{pages}{1157--1166}, \doiprefix\url{https://doi.org/10.1002/jcc.23910} (\bibinfo{year}{2015}).
\newblock \eprint{https://onlinelibrary.wiley.com/doi/pdf/10.1002/jcc.23910}.

\bibitem{MACE-OFF23}
\bibinfo{author}{Kovács, D.~P.} \emph{et~al.}
\newblock \bibinfo{title}{Mace-off23: Transferable machine learning force fields for organic molecules} (\bibinfo{year}{2023}).
\newblock \eprint{2312.15211}.

\bibitem{DPA-2}
\bibinfo{author}{Zhang, D.} \emph{et~al.}
\newblock \bibinfo{title}{Dpa-2: a large atomic model as a multi-task learner} (\bibinfo{year}{2024}).
\newblock \eprint{2312.15492}.

\bibitem{M2Hub}
\bibinfo{author}{Du, Y.} \emph{et~al.}
\newblock \bibinfo{title}{M\^{}2hub: Unlocking the potential of machine learning for materials discovery}.
\newblock In \bibinfo{editor}{Oh, A.} \emph{et~al.} (eds.) \emph{\bibinfo{booktitle}{Advances in Neural Information Processing Systems}}, vol.~\bibinfo{volume}{36}, \bibinfo{pages}{77359--77378} (\bibinfo{publisher}{Curran Associates, Inc.}, \bibinfo{year}{2023}).

\bibitem{Zhao2021}
\bibinfo{author}{Zhao, Q.} \& \bibinfo{author}{Savoie, B.~M.}
\newblock \bibinfo{journal}{\bibinfo{title}{Simultaneously improving reaction coverage and computational cost in automated reaction prediction tasks}}.
\newblock {\emph{\JournalTitle{Nature Computational Science}}} \textbf{\bibinfo{volume}{1}}, \bibinfo{pages}{479--490}, \doiprefix\url{10.1038/s43588-021-00101-3} (\bibinfo{year}{2021}).

\bibitem{Doobs2024}
\bibinfo{author}{Du, Y.} \emph{et~al.}
\newblock \bibinfo{title}{Doob's lagrangian: A sample-efficient variational approach to transition path sampling} (\bibinfo{year}{2024}).
\newblock \eprint{2410.07974}.

\bibitem{equivariance}
\bibinfo{author}{Serre, J.-P.} \emph{et~al.}
\newblock \emph{\bibinfo{title}{Linear representations of finite groups}}, vol.~\bibinfo{volume}{42} (\bibinfo{publisher}{Springer}, \bibinfo{year}{1977}).

\bibitem{gdlbook}
\bibinfo{author}{Bronstein, M.~M.}, \bibinfo{author}{Bruna, J.}, \bibinfo{author}{Cohen, T.} \& \bibinfo{author}{Veli{\v{c}}kovi{\'c}, P.}
\newblock \bibinfo{journal}{\bibinfo{title}{Geometric deep learning: Grids, groups, graphs, geodesics, and gauges}}.
\newblock {\emph{\JournalTitle{arXiv:2104.13478}}}  (\bibinfo{year}{2021}).

\bibitem{frank}
\bibinfo{author}{K{\"o}hler, J.}, \bibinfo{author}{Klein, L.} \& \bibinfo{author}{No{\'e}, F.}
\newblock \bibinfo{title}{Equivariant flows: {E}xact likelihood generative learning for symmetric densities}.
\newblock In \emph{\bibinfo{booktitle}{International Conference on Machine Learning}}, \bibinfo{pages}{5361--5370} (\bibinfo{year}{2020}).

\bibitem{villani2009optimal}
\bibinfo{author}{Villani, C.} \emph{et~al.}
\newblock \emph{\bibinfo{title}{Optimal transport: old and new}}, vol. \bibinfo{volume}{338} (\bibinfo{publisher}{Springer}, \bibinfo{year}{2009}).

\bibitem{villani2021topics}
\bibinfo{author}{Villani, C.}
\newblock \emph{\bibinfo{title}{Topics in optimal transportation}}, vol.~\bibinfo{volume}{58} (\bibinfo{publisher}{American Mathematical Soc.}, \bibinfo{year}{2021}).

\bibitem{santambrogio2015optimal}
\bibinfo{author}{Santambrogio, F.}
\newblock \bibinfo{journal}{\bibinfo{title}{Optimal transport for applied mathematicians}}.
\newblock {\emph{\JournalTitle{Birk{\"a}user, NY}}} \textbf{\bibinfo{volume}{55}}, \bibinfo{pages}{94} (\bibinfo{year}{2015}).

\bibitem{Linfeng2018MPFlow}
\bibinfo{author}{Zhang, L.}, \bibinfo{author}{E, W.} \& \bibinfo{author}{Wang, L.}
\newblock \bibinfo{title}{Monge-amp\`ere flow for generative modeling} (\bibinfo{year}{2018}).
\newblock \eprint{1809.10188}.

\bibitem{monge1781memoire}
\bibinfo{author}{Monge, G.}
\newblock \bibinfo{journal}{\bibinfo{title}{M{\'e}moire sur la th{\'e}orie des d{\'e}blais et des remblais}}.
\newblock {\emph{\JournalTitle{Mem. Math. Phys. Acad. Royale Sci.}}} \bibinfo{pages}{666--704} (\bibinfo{year}{1781}).

\bibitem{kantorovich1942translocation}
\bibinfo{author}{Kantorovich, L.~V.}
\newblock \bibinfo{title}{On the translocation of masses}.
\newblock In \emph{\bibinfo{booktitle}{Dokl. Akad. Nauk. USSR (NS)}}, vol.~\bibinfo{volume}{37}, \bibinfo{pages}{199--201} (\bibinfo{year}{1942}).

\bibitem{benamou2000computational}
\bibinfo{author}{Benamou, J.-D.} \& \bibinfo{author}{Brenier, Y.}
\newblock \bibinfo{journal}{\bibinfo{title}{A computational fluid mechanics solution to the monge-kantorovich mass transfer problem}}.
\newblock {\emph{\JournalTitle{Numerische Mathematik}}} \textbf{\bibinfo{volume}{84}}, \bibinfo{pages}{375--393} (\bibinfo{year}{2000}).

\bibitem{peyre2017computational}
\bibinfo{author}{Peyr{\'e}, G.} \& \bibinfo{author}{Cuturi, M.}
\newblock \bibinfo{journal}{\bibinfo{title}{{Computational optimal transport}}}.
\newblock {\emph{\JournalTitle{Center for Research in Economics and Statistics Working Papers}}}  (\bibinfo{year}{2017}).

\bibitem{liu2022flow}
\bibinfo{author}{Liu, X.}, \bibinfo{author}{Gong, C.} \& \bibinfo{author}{Liu, Q.}
\newblock \bibinfo{journal}{\bibinfo{title}{Flow straight and fast: Learning to generate and transfer data with rectified flow}}.
\newblock {\emph{\JournalTitle{arXiv preprint arXiv:2209.03003}}}  (\bibinfo{year}{2022}).

\bibitem{shen2021accurate}
\bibinfo{author}{Shen, Z.} \emph{et~al.}
\newblock \bibinfo{journal}{\bibinfo{title}{Accurate point cloud registration with robust optimal transport}}.
\newblock {\emph{\JournalTitle{Advances in Neural Information Processing Systems}}} \textbf{\bibinfo{volume}{34}}, \bibinfo{pages}{5373--5389} (\bibinfo{year}{2021}).

\bibitem{titouan2019optimal}
\bibinfo{author}{Titouan, V.}, \bibinfo{author}{Courty, N.}, \bibinfo{author}{Tavenard, R.} \& \bibinfo{author}{Flamary, R.}
\newblock \bibinfo{title}{Optimal transport for structured data with application on graphs}.
\newblock In \emph{\bibinfo{booktitle}{International Conference on Machine Learning}}, \bibinfo{pages}{6275--6284} (\bibinfo{organization}{PMLR}, \bibinfo{year}{2019}).

\bibitem{Repaint}
\bibinfo{author}{Lugmayr, A.} \emph{et~al.}
\newblock \bibinfo{title}{Repaint: Inpainting using denoising diffusion probabilistic models}.
\newblock In \emph{\bibinfo{booktitle}{2022 IEEE/CVF Conference on Computer Vision and Pattern Recognition (CVPR)}}, \doiprefix\url{10.1109/CVPR52688.2022.01117} (\bibinfo{year}{2022}).

\end{thebibliography}

\clearpage

\setstretch{1}

\appendix

\title{\textit{Supplementary Information} for "Optimal Transport for Generating Transition State in Chemical Reactions"}

\maketitle

\renewcommand{\thesection}{\arabic{section}}  
\renewcommand{\thetable}{\arabic{table}}  
\renewcommand{\thefigure}{\arabic{figure}}
\setcounter{figure}{0}
\setcounter{table}{0}

\makeatletter
\renewcommand{\fnum@figure}{\textbf{Figure \thefigure}. }
\renewcommand{\fnum@table}{\textbf{Table \thetable. }}


\addcontentsline{toc}{section}{Abbreviation}
\section*{Abbreviation}
The following is the list of abbreviation utilized in the main paper.
\begin{enumerate}
    \item React-OT: \underline{O}ptimal \underline{t}ransport for elementary \underline{react}ions
    \item OA-ReactDiff: \underline{O}bject-\underline{a}ware SE(3) GNN for generating sets of 3D molecules in elementary \underline{react}ions under the \underline{diff}usion model
    \item RMSD: Root mean square deviation.
    \item SE(3): Special Euclidean group in 3D space.
    \item TS: Transition state.
    \item MAE: Mean absolute error.
\end{enumerate}

\section{Physical symmetries and constraints in an elementary reaction.}
\label{Supp:required_symmetries}
An elementary reaction that consists of $n$ fragments as reactant and $m$ fragments as product can be described as $\{\mathrm{R}^{(1)}, ..., \mathrm{R}^{(n)}, \mathrm{TS}, \mathrm{P^{(1)}}, ...,  \mathrm{P^{(m)}}\}$. This reaction requires the following symmetries:
\begin{enumerate}
    \item \textit{Permutation symmetry among atoms in a fragment}. For any fragment in $\mathrm{R}^{(i)}, \mathrm{TS}, \mathrm{P^{(j)}}$, change of atom ordering preserves the reaction.
    \item \textit{Permutation symmetry among fragments in reactant and product}. The change of ordering in $\{\mathrm{R}^{(1)}, ..., \mathrm{R}^{(n)} \}$ and $\{\mathrm{P}^{(1)}, ..., \mathrm{P}^{(m)} \}$ preserve the reaction.
    \item \textit{Rotation and translation symmetry for each fragment}. Rotation and translation operations on any fragment (i.e., $\mathrm{R}^{(i)}, \mathrm{TS}, \mathrm{P^{(j)}}$) preserve the reaction.
\end{enumerate}

\section{Reaction network exploration with React-OT.}
\label{Supp:reaction_network}
To demonstrate the practical application of React-OT in reaction network exploration, the reaction network of $\gamma$-ketohydroperoxide (KHP), a well-studied system commonly used as a benchmark in recent studies.\cite{grambow2018KHP,naz2020unimolecular,QiyuanNCS2021,zhao2022YARP2} Building on reactions generated via graph-based enumeration,\cite{QiyuanNCS2021} we utilized React-OT to determine transition states, which are then used to evaluate activation energies at $\omega$B97X/6-31G* level of theory. A two-step reaction network was constructed based on the predicted activation energies and compared with the network generated by the Yet Another Reaction Program (YARP).\cite{zhao2022YARP2} To manage network size and emphasize kinetically significant reactions, a straightforward growth rule was applied: each node expansion permits the generation of up to five branches, prioritizing reactions with the lowest energy barriers. 

\begin{figure*}[t!]
    \includegraphics[width=0.7\textwidth]{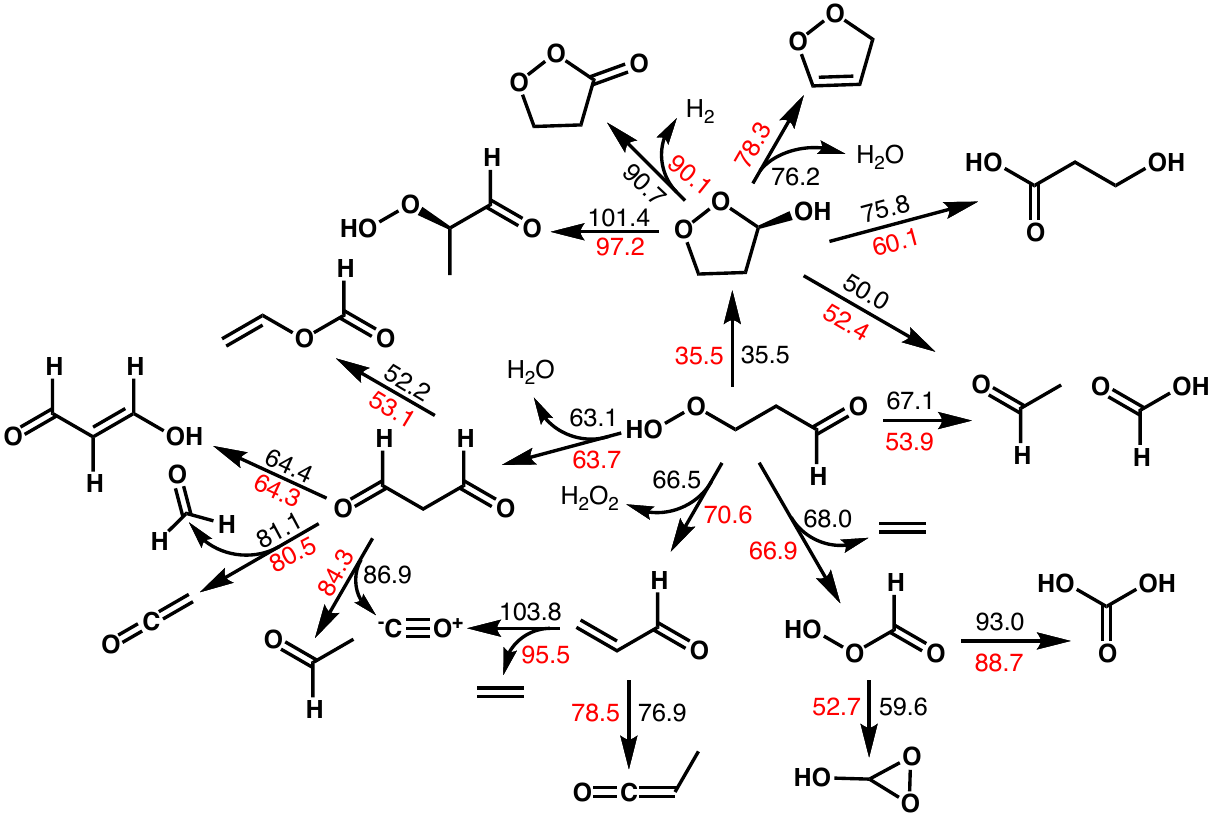}
    \caption{\textbf{Reaction Network of $\gamma$-ketohydroperoxide (KHP)}. A two-step reaction network of KHP generated by the Yet Another Reaction Program. Numbers denoted by red/black refer to activation energies computed on DFT-optimized/React-OT-generated transition states.}
    \label{Supp:KHPNET}
\end{figure*}

The resulting network generated by React-OT has identical nodes as the one generated by YARP (note: reactant and transition state energies are evaluated at $\omega$B97X/6-31G*, and B3LYP-D3/TZVP//$\omega$B97X/6-31G* level of theory, respectively), indicating that all key reactions were captured, with a mean absolute energy difference of 3.84 kcal/mol (Fig. \ref{Supp:KHPNET}). This case study illustrates how React-OT can be applied to speed up the reaction network exploration.

\begin{figure*}[htbp]
    \includegraphics[width=0.58\textwidth]{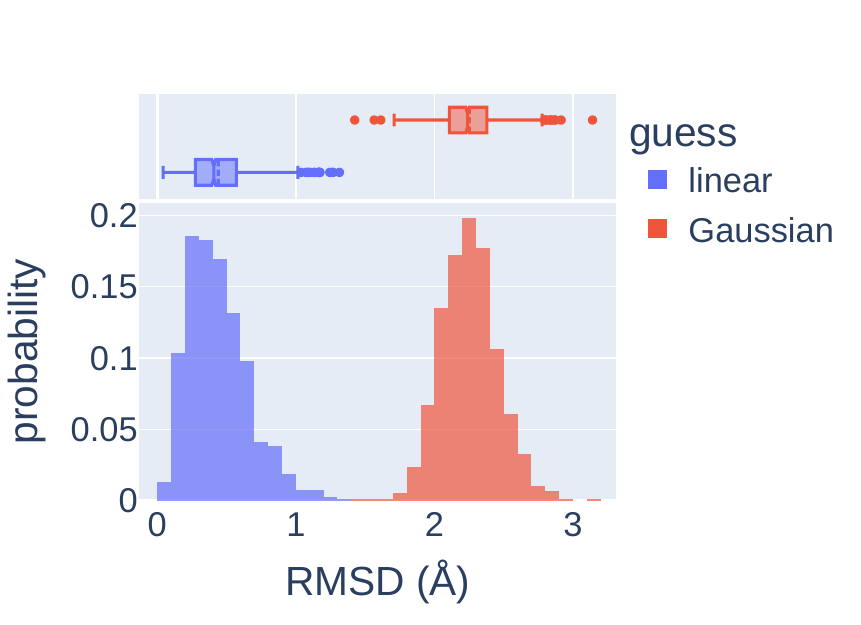}
    \caption{\textbf{RMSD between the true TS different initial guesses.} 
    Linear interpolation between reactants and products (blue) and samples from a Gaussian distribution (red).
    }
    \label{Supp:rmsd_guss}
\end{figure*}

\begin{figure*}[t!]
    \includegraphics[width=0.68\textwidth]{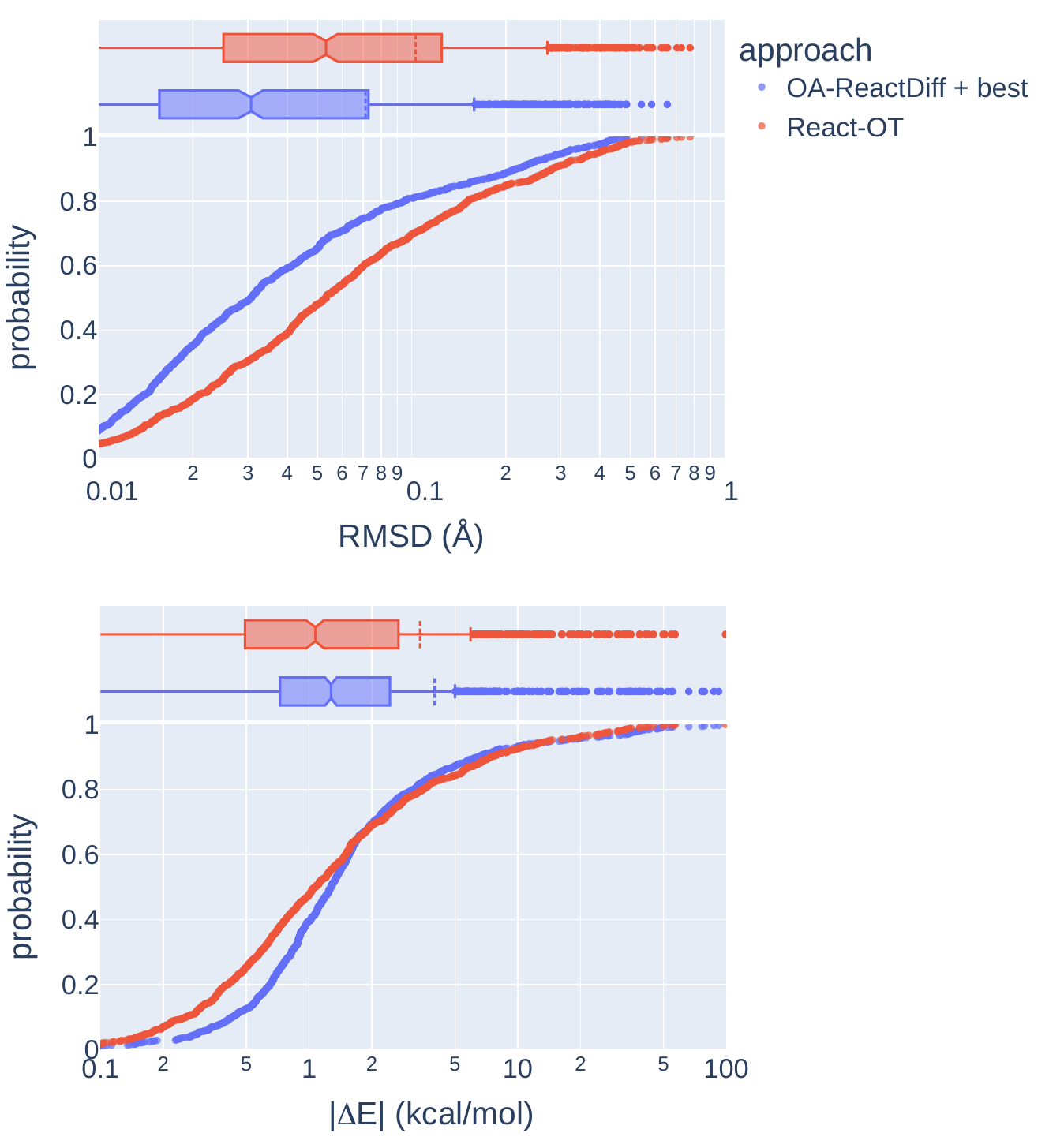}
    \caption{\textbf{Additional performance comparison between React-OT and OA-ReactDiff.}
    Cumulative probability for RMSD (top) and barrier height (bottom) for React-OT (red) and OA-ReactDiff + best sample in 40 runs (blue) both shown in log scale for visibility.
    Note that the performance of OA-ReactDiff + best sample in 40 runs is not practically achievable as one does not know the best sample a priori.
    }
    \label{Supp:ot_vs_diff_best}
\end{figure*}

\begin{figure*}[t!]
    \includegraphics[width=0.6\textwidth]{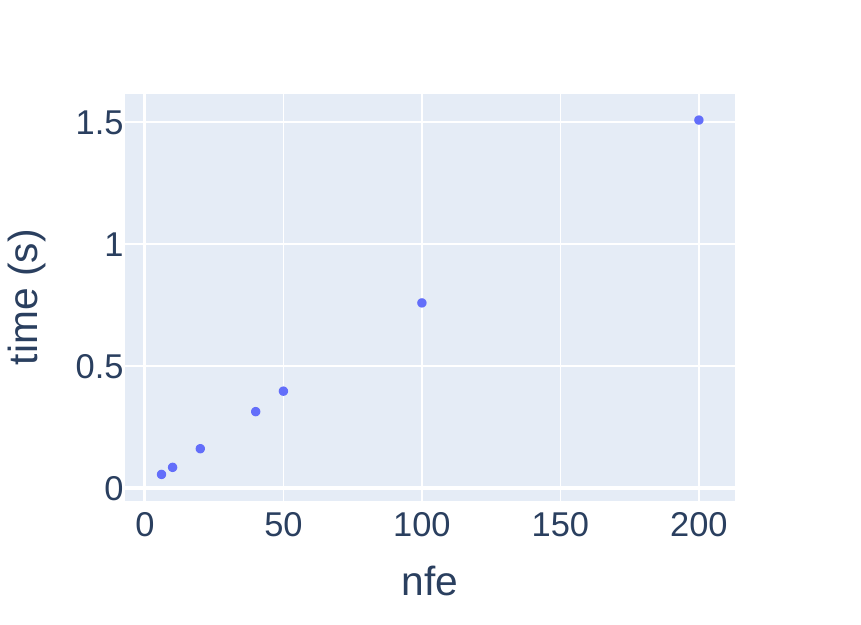}
    \caption{\textbf{Sampling cost at different number of function evaluation.}
    }
    \label{Supp:nfe_vs_time}
\end{figure*}

\begin{figure*}[t!]
    \includegraphics[width=0.55\textwidth]{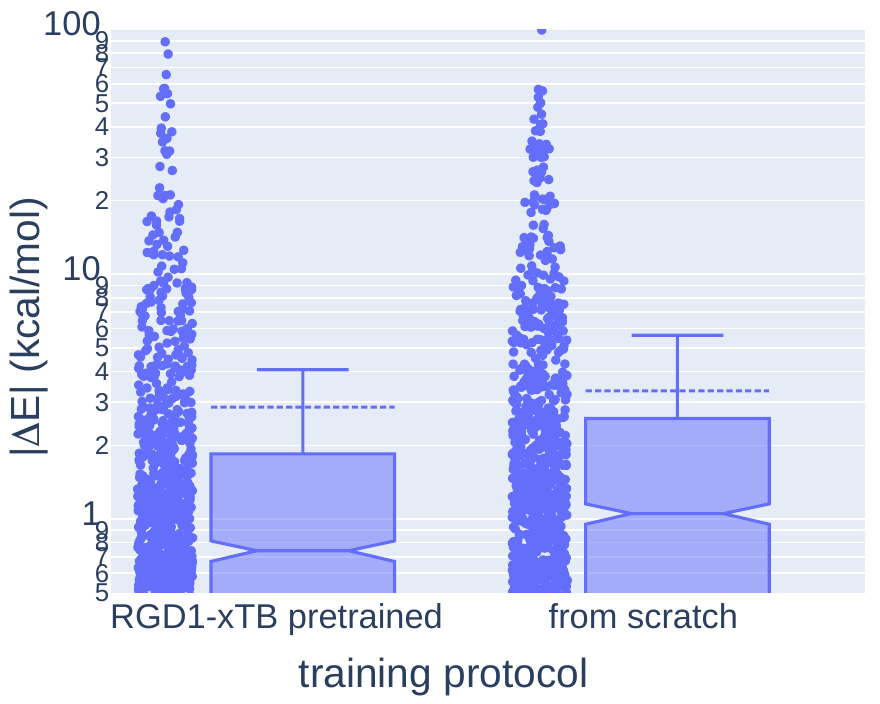}
    \caption{\textbf{Persistent barrier height errors.}
    Box plots with all points shown for barrier height errors obtained by React-OT with RGD1-xTB pretraining and training from scratch. 
    The barrier height is shown in log scale with a trucation of 0.5 kcal/mol.
    }
    \label{Supp:persistent_high_error}
\end{figure*}

\begin{figure*}[t!]
    \includegraphics[width=0.5\textwidth]{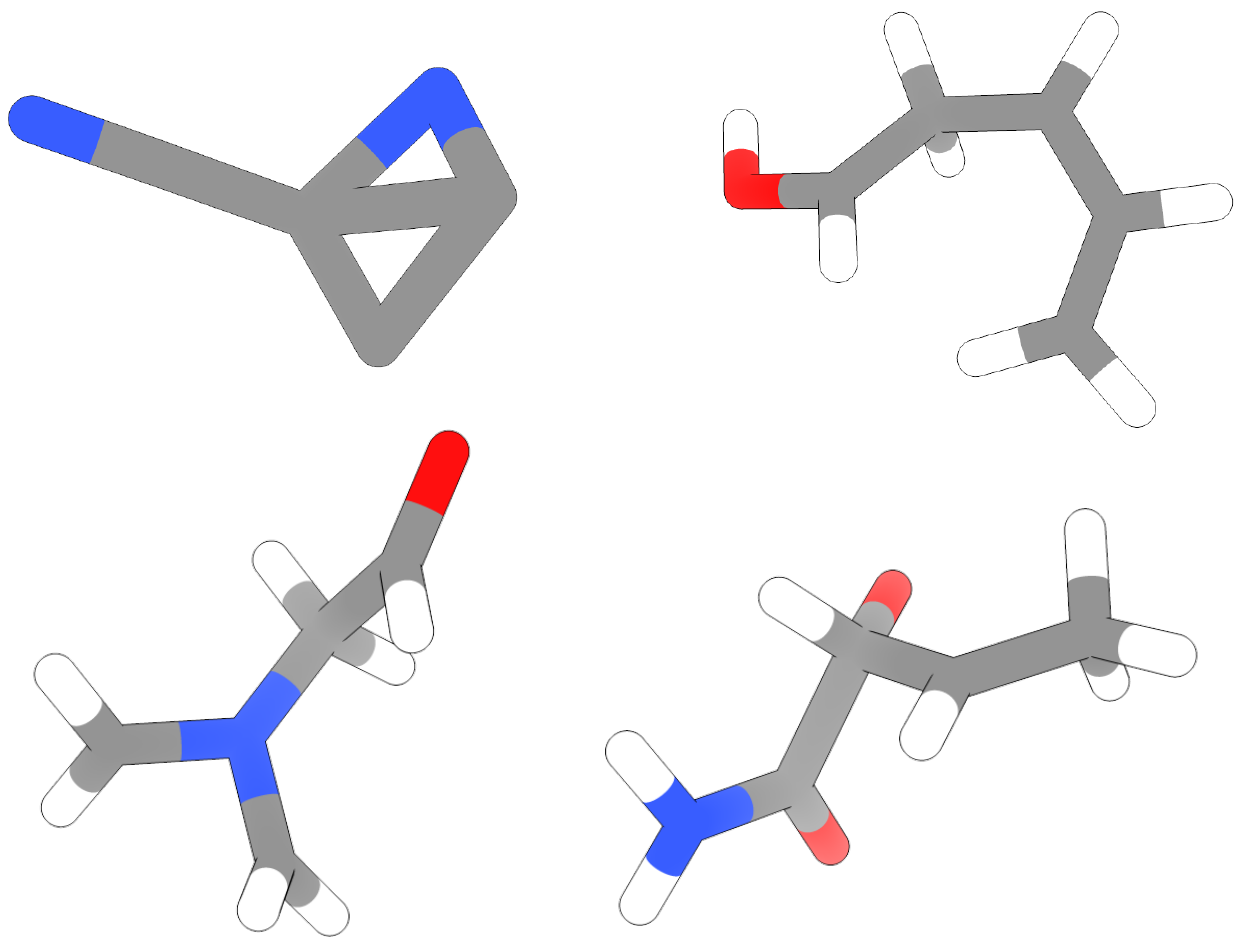}
    \caption{\textbf{Examples of TS structures breaking the octet rule that React-OT gives large errors.}
    Atoms colored as follows: gray for C, blue for N, red for O, and white for H.
    }
    \label{Supp:non_octect_examples}
\end{figure*}

\begin{figure*}[t!]
    \includegraphics[width=0.5\textwidth]{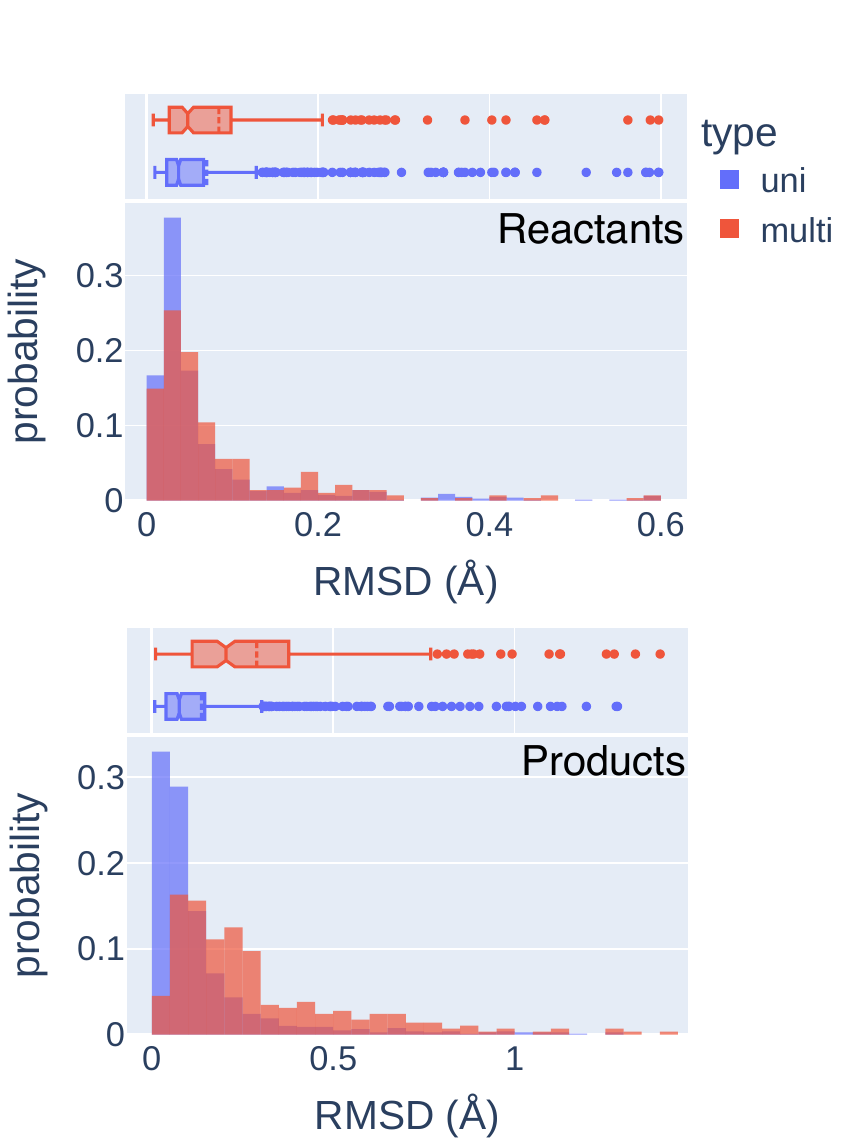}
    \caption{\textbf{RMSD between xTB and DFT optimized reactants and products.}
    Distribution of RMSD for reactants (top) and products (bottom) grouped by the reaction type: uni-molecular reaction in blue and multi-molecular reaction in red.
    }
    \label{Supp:xtb_rmsd}
\end{figure*}

\begin{figure*}[t!]
    \includegraphics[width=0.55\textwidth]{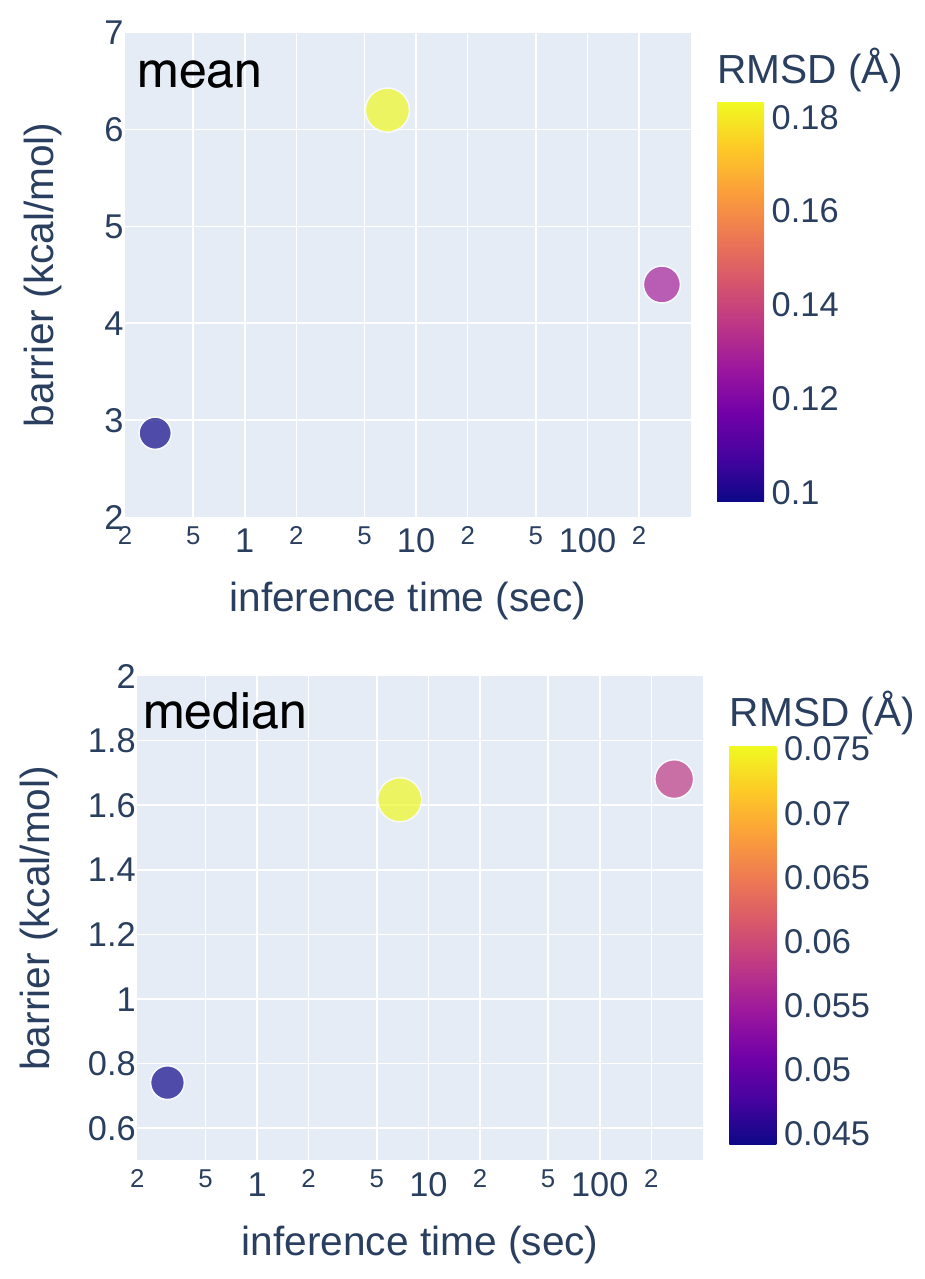}
    \caption{\textbf{Summary of performance for React-OT and OA-ReactDiff}.
    Barrier height, RMSD (mean as top and median as bottom), and inference time for React-OT (blue), one-shot OA-ReactDiff (yellow), and 40-shot OA-ReactDiff with recommender (purple).
    }
    \label{Supp:sumamry}
\end{figure*}

\begin{figure*}[t!]
    \includegraphics[width=0.6\textwidth]{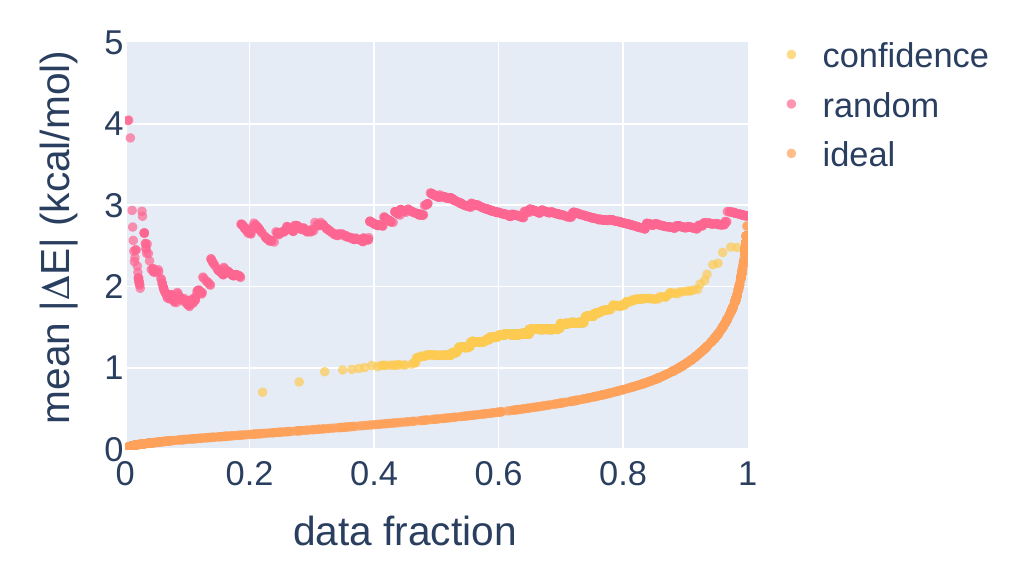}
    \caption{\textbf{Mean barrier height of React-OT with different methods for adding confidence data.}
    Yellow corresponds to the confidence model used in the main text.
    Pink corresponds to random selection, giving mean errors oscillating around the overall model performance.
    Orange corresponds to the ideal case, where the lowest error data point is always selected as the data fraction increases.
    }
    \label{Supp:random_confidence}
\end{figure*}

\begin{table*}[th]
\centering 
\caption{\textbf{Statistics of RMSD between xTB and DFT optimized reactants and products.}.
}
\resizebox{0.4\textwidth}{!}{
\begin{tabular}{l|c|cc}\toprule
\multicolumn{1}{c|}{Species} & \multicolumn{1}{c|}{Reaction type} &\multicolumn{2}{c}{RMSD (Å)} \\\midrule
& &mean &median\\\midrule
reactant&uni-molecular&0.070&0.037\\
reactant&multi-molecular&0.084&0.048\\
product&uni-molecular&0.137&0.076\\
product&multi-molecular&0.289&0.204\\
\bottomrule
\end{tabular}}
\label{Supp:table_xtb_rmsd}
\end{table*}

\begin{figure*}[t!]
    \includegraphics[width=0.5\textwidth]{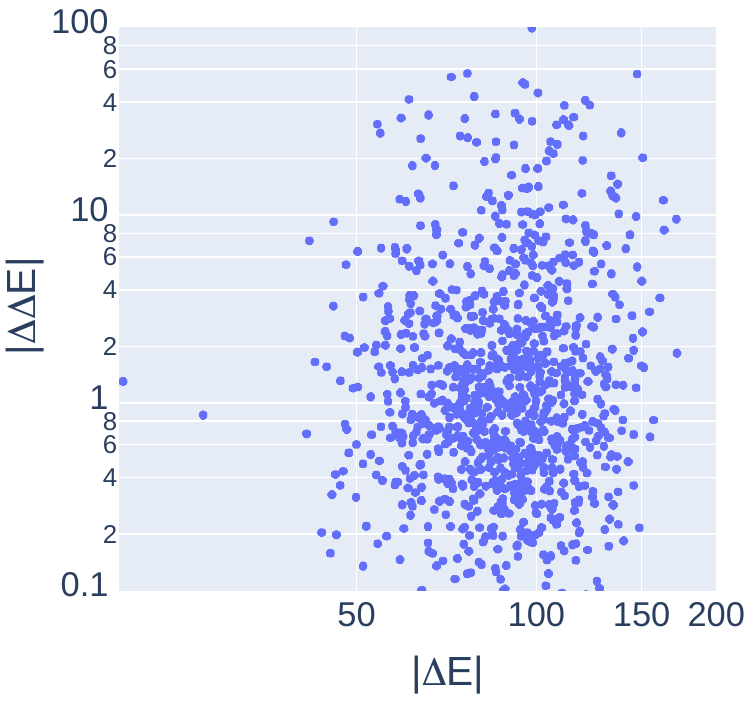}
    \caption{\textbf{Actual vs. React-OT error of barrier height on Transition1x test data.}
    }
    \label{Supp:abs_barrier_vs_barrier_error}
\end{figure*}

\begin{figure*}[t!]
    \includegraphics[width=0.5\textwidth]{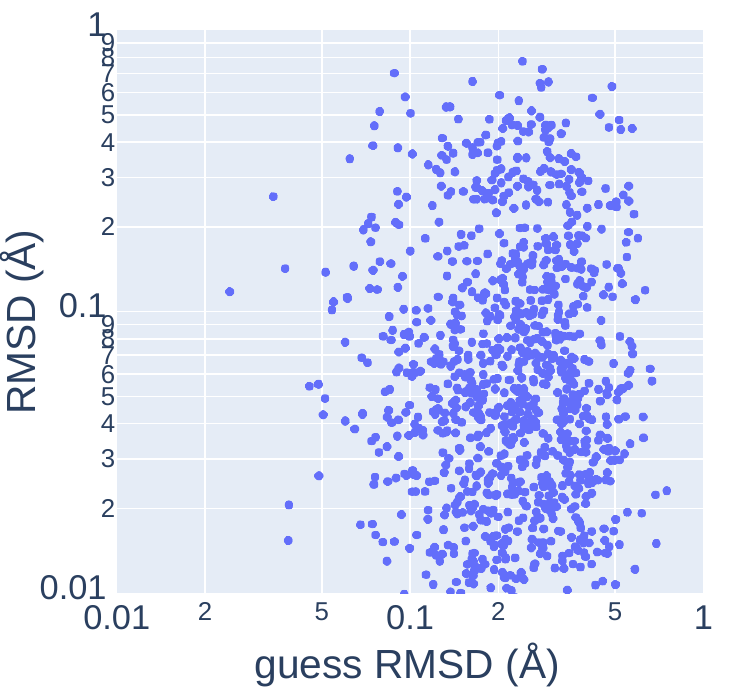}
    \caption{\textbf{RMSD of initial guess vs. React-OT generated TS on Transition1x test data.}
    }
    \label{Supp:guess_vs_ot_rmsd}
\end{figure*}

\begin{figure*}[t!]
    \includegraphics[width=0.5\textwidth]{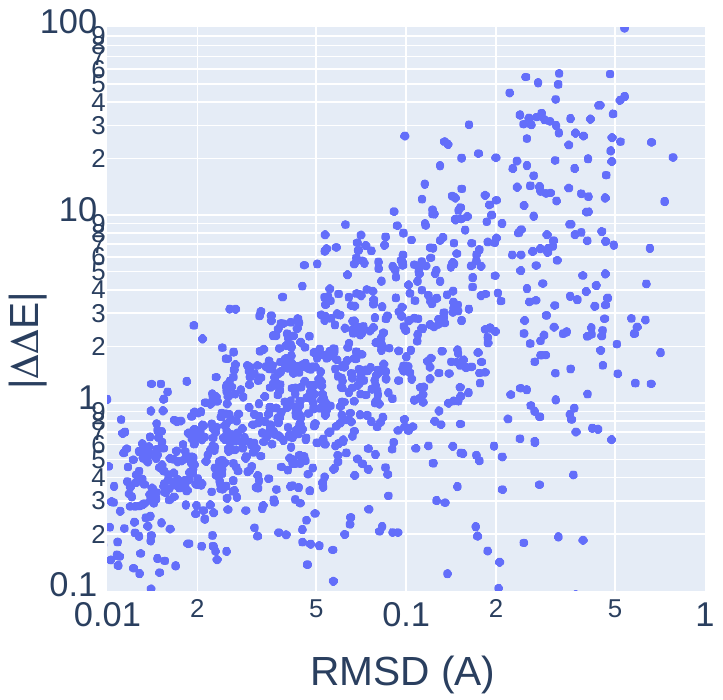}
    \caption{\textbf{RMSD vs. barrier heigh error for React-OT generated TS on Transition1x test data.}
    }
    \label{Supp:rmsd_vs_barrier_error}
\end{figure*}

\begin{figure*}[t!]
    \includegraphics[width=0.5\textwidth]{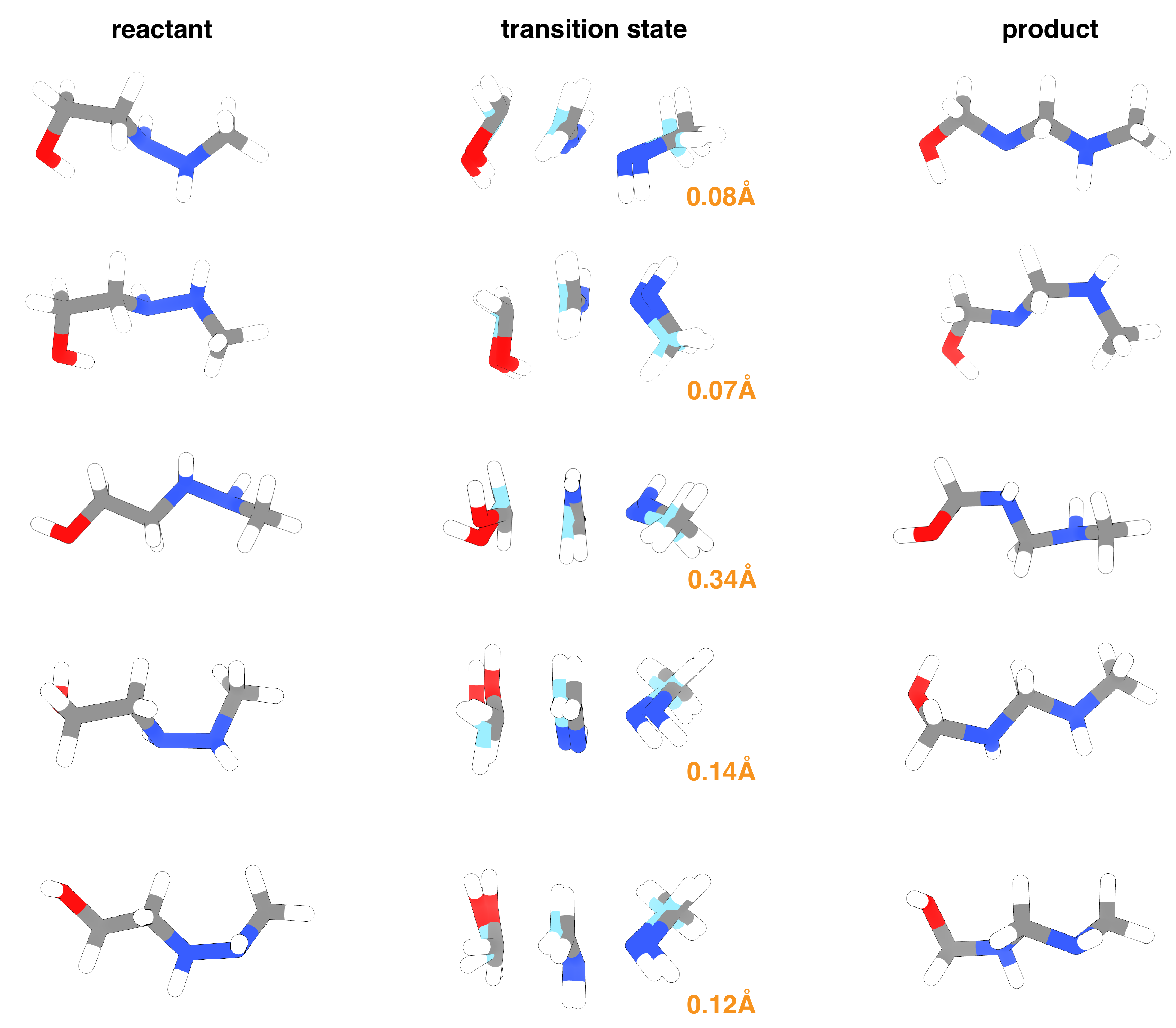}
    \caption{\textbf{TS structures generated by React-OT at different reactant and product conformations.}
    React-OT applied on a self-rearrangement reaction at five different conformations, where all reactants (left), TS (middle) and products (right) are shown. For TS, React-OT generated structures are colored with light blue in their C atoms, while the true TS structures are colored with gray in their C atoms. RMSD for each TS conformation is shown in orange.
    }
    \label{Supp:reaction_conformations}
\end{figure*}

\begin{table*}[th]
\centering 
\caption{\textbf{Ablation study on React-OT}.
}
\resizebox{0.4\textwidth}{!}{
\begin{tabular}{l|cc}\toprule
\multicolumn{1}{c|}{Model} &\multicolumn{2}{c}{RMSD (Å)} \\\midrule
&mean &median\\\midrule
current&0.1029&0.0527\\
w/o diffusion training&0.1632&0.1202\\
initial TS from normal distribution&0.9860&> 1.0\\
initial TS as reactants&0.1989&0.0962\\
\bottomrule
\end{tabular}}
\label{Supp:table_ablation_study}
\end{table*}

\begin{table*}[th]
\centering 
\caption{\textbf{Transferability of React-OT on out-of-distribution datasets}.
}
\resizebox{0.4\textwidth}{!}{
\begin{tabular}{l|cc}\toprule
\multicolumn{1}{c|}{Dataset} &\multicolumn{2}{c}{RMSD (Å)} \\\midrule
&mean &median\\\midrule
Transition1x&0.1029&0.0527\\
Berkholz-15&0.1152&0.0503\\
DielsAlde-41&0.0495&0.0402\\
\bottomrule
\end{tabular}}
\label{Supp:table_ablation_study}
\end{table*}

\end{document}